\documentclass[aps,pre,10pt,floatfix,nofootinbib]{revtex4-2}
\usepackage{graphicx}
\usepackage{amsmath}

\usepackage{amssymb}

\usepackage{capt-of}

\usepackage[section]{placeins}

\usepackage{multirow}

\usepackage{natbib}

\numberwithin{table}{section}

\usepackage{enumitem}

\usepackage{array}   
\newcolumntype{L}{>{$}l<{$}} 
\newcolumntype{C}{>{$}c<{$}}
\newcolumntype{R}{>{$}r<{$}}

\usepackage{listings}

\begin{document}

\title{Using thymine-18 for enhancing dose delivery and localizing the Bragg peak in proton-beam therapy}

\author{William Parke}
\affiliation{The George Washington University, Washington, DC \\}

\author{Dalong Pang}
\affiliation{Georgetown Medical School, Washington, DC\\}

\date{\today\  v0.7}

\begin{abstract}

Therapeutic protons acting on O18-substituted thymidine
increase cytotoxicity in radio-resistant human cancer cells.
We consider here the physics behind the irradiation during proton beam therapy and diagnosis using O18-enriched thymine in DNA, 
with attention to the effect of the presence of thymine-18 on cancer cell death.

\end{abstract}

\maketitle

\section{introduction}

This technical report presents the physics background behind a proposal for proton-beam therapy (PT) using doped DNA in which one of the oxygen atoms in the thymine bases have been replaced by oxygen-18 (a stable isotope), for the purpose of enhanced dose to tumors and for the side benefit of localization of
the peak-dose delivery region.

\begin{figure}[thpb]
\caption{Thymine-18. $\,^{18}$O is attached to the carbon between the two nitrogens. (This carbon is
atom \#2 in the ring, with the second oxygen attached to atom \#4, and the methyl group attached to atom \#5, also carbon atoms.)}
\center
\includegraphics[width=0.3\columnwidth]{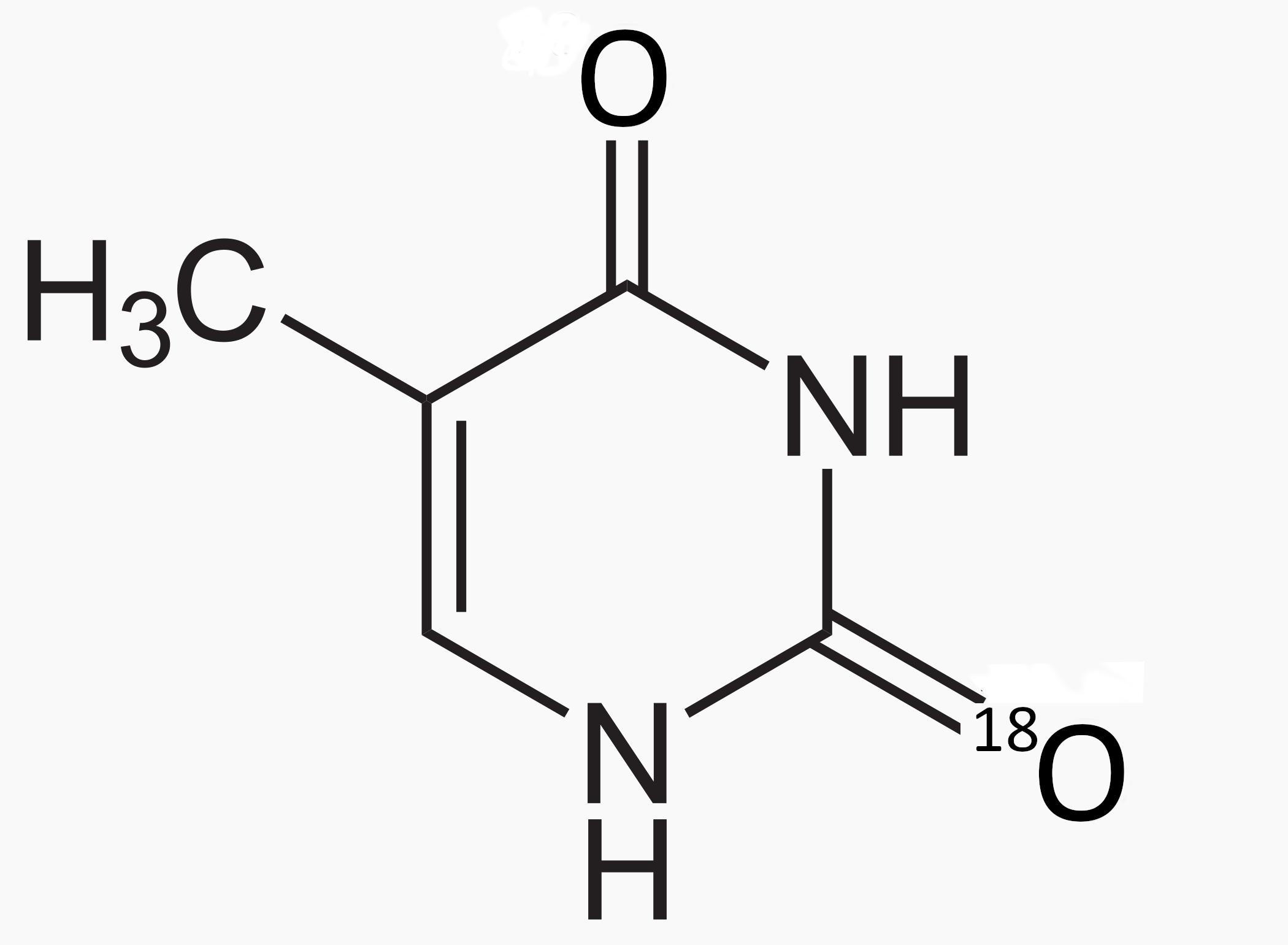}
\label{fig-0}
\end{figure}

Proton therapy delivers radiation to cancers to cause malignant-cell death, principally by  radical formation causing unrepairable damage to their DNA and thereby slowing or stopping tumor growth. Therapeutic protons, in passing through tissue, will scatter electrons and generate ions and radicals, such as hydroxyl. Through chemical reactions, the radicals can cause single-strand breaks (SSB) in DNA.  Two such nearby breaks can generate a double-strand break (DSB). With a
much lower probability, a direct DSB can be caused by a beam proton passing through the DNA helix itself,
producing a `clustered' pair of SSBs. Beam protons can also activate nuclei, such as those of oxygen, carbon and nitrogen atoms.
These, in turn, can release, through strong-interaction decay, gammas and fast-moving secondary particles such as protons, neutrons, deuterons, tritium and alpha particles, and through weak-interaction decay, electrons, positrons, and neutrinos. Those secondaries with a charge can also produce localized damage to
the functioning of tumor cells.

A recent project \cite{rich2020}) at the Georgetown University Hospital initiated a study of the lethality enhancement of irradiated cells due to modified DNA, replacing the O16 with  O18 in thymine (see Figure \ref{fig-0}). Upon proton irradiation, some O18 is transformed to F18, producing a DNA breakage mechanism and a positron-emission tomography (PET) signal,
which can be monitored to help track where the proton beam Bragg peak occurred in the doped tissue. The Georgetown group exposed SQ20-B squamous carcinoma cells to physiologic $\,^{18}$O-thymidine concentration of 5 $\mu$M for 48 hours followed by $1$ to $9$ Gy graded doses of proton radiation given $24$ hours later. Survival analysis showed a dose modification factor ($DMF$) of 1.2 due to the substituted thymidine.

Here we wish to estimate the rate of  O18 conversion to F18 for a given proton beam and fixed amount of doped DNA, and the subsequent resulting DNA breakage induced. Also, we want to assess the use of the PT-produced F18 to better localize the PT beam region of maximum dose via a post-PT PET scanning.

\section{Physics Analysis}

\subsection{Proton beams in tissue}

When an energetic proton (initial kinetic energy $K_0$ from 50 to 250 MeV) enters tissue, the atoms and molecules in the tissue will slow the beam protons, largely through scattering and ionization of
tissue electrons.  Because the proton is 1836 times more massive than the electron,
the direction of the proton is not significantly changed by electron encounters. (Appendix \ref{app-max-theta} shows that for each encounter of a proton with a quasi-free electron, the proton deviates from a straight line no more than $\arcsin{(m_e/m_p)}=0.0312\,$deg, where $m_e$ is the mass of the
electron and $m_p$ is the mass of the proton.)
The scattered electrons (`delta' electrons) spread the beam ionization over nearby tissue, but not
more than about $2$ mm for 200 MeV protons.  The `stopping power' (loss of beam-proton kinetic energy with distance, $-dK/dz$) depends on the density of  the tissue, largely because of the electron density. Fluences in the range of $10^9$ protons per square centimeter are typical for proton-beam therapy. The fluence rate $d\Phi/dt$ of protons in the beam decreases as protons are
taken from the beam, with interaction of beam protons with nuclei in the tissue a large contributor to the fluence loss. These nuclear events are relatively rare
because of the small size of the nuclei, but can widely deflect beam protons by nuclear-Coulomb
elastic and inelastic scattering. Nonelastic nuclear reactions also occur when the beam protons
have sufficient energy to penetrate the nuclear Coulomb barrier, making
fast-moving reaction products, generally moving away from the proton beam axis.%
\footnote{We will take elastic collision to mean that the incoming and outgoing particles are the same, and that all the initial kinetic energy is returned to the outgoing particles.  A quasi-elastic collision occurs if the incoming particle knocks out a particle in a bound state of a target, with little energy being transferred to the other particles in the target. In an inelastic collision, some of the initial kinetic energy is converted into internal energy in one or more of the outgoing particles. For a noneleastic collision, the set of outgoing particles differs in internal structure from the incoming ones.}

The physical dose delivered by the beam within some small volume $dV$ of tissue, i.e. the energy $dE_{dep}$ deposited in that volume per unit mass of tissue
\begin{equation}
D=\frac{dE_{dep}}{dm}=\frac{1}{\rho}\frac{dE_{dep}}{dV} \ ,
\end{equation}
comes from electron ionizations (producing chemical transformations and radical formation), but is also affected by the energy deposited by the products of nuclear reactions. For a proton beam with an initial energy of $90\,$MeV, $5\,$\% of the beam energy loss from the beam comes from inelastic nuclear reactions (\cite{seltzer1993}).

The `biologically effective dose' ($BED$), $D_{BE}$, with units sievert (Sv), defined as the physical dose times a unitless factor called the `relative biological effectiveness' ($RBE$), attempts to make the $BED$ have the same biologically damaging effect as photons with about the same energy. The overall $RBE$ of proton beams is around $1.1$, while for carbon ions, its about $2.5$ and for neutrons $RBE$ exceeds $3$. However, the $RBE$ of a proton beam depends on its energy, reaching values as much as 4 or 5 when the beam's rate of energy loss is about 100 keV per micrometer (\cite{seltzer1993}) because of a larger number of atomic electron excitation and ionization, but also having contributions from a greater number proton-nuclei reactions making ionizing secondaries.  A $BED$ over $1\,$Sv will frequently be fatal to living cells (lethality is $50\,$\% with a $BED$ of 4 to 5 Sv delivered over a few minutes).

In tissue, the released neutrons are likely to bounce around among the nearby nuclei until absorbed by a nucleus or thermalized. The energy loss is slow relative to protons, because there is no Coulomb interaction, so the neutron must get within about a fermi of a nucleus to scatter. For neutrons in the 1 to 20 MeV energy range, scattering against a high Z nucleus can cause that nucleus to be excited, but this has a lower probability than elastic scattering. At lower energies, the neutron is likely to be absorbed, even by hydrogen protons making deuterium.  The neutron attenuation coefficient depends on the inverse of its relative speed for energies below about 1 eV.  The time for diffusive thermalization ranges from about 200 microseconds (water) to 20 milliseconds (graphite). A thermalized neutron is then likely to be absorbed by a nearby nucleus.  Without a nuclear-absorption event, the neutron will beta decay (with a half-life time of $10.2\,$min) to a proton, electron, and anti-neutrino, those products sharing $0.782\,$MeV in kinetic energy. 
The neutrinos exit with practically no interaction
with matter on their path. (A $1\,$MeV electron anti-neutrino has a mean-free path 
through water of $60\,$light-years!)

The fluence of secondary neutrons produced by proton beams in tissue is low \cite{cascio2005}, but because these neutrons are easily spread, and their relative biological effectiveness can be 10 to 20 times higher than the protons, their presence cannot be ignored.  However, with proper strategies, the neutron-generated dose to healthy tissue 
is still low with proton beam therapy when compared to the dose to healthy tissue under photon beam therapy\cite{schneider2002}.

\subsection{Using Oxygen-18}

Among the many reactions of an energetic proton-beam with a thymine-18 doped target, the reaction  \[\,^{18}\text{O}(p,n)\,^{18}\text{F}\] has a relatively large cross section (hundreds of millibarns)
in the $4$ to $14\,$MeV proton kinetic energy range.
The presence  of this reaction can be monitored because of the subsequent beta decay of fluorine-18. The beta decay produces a positron, which may scatter with local nuclei and electrons, and then annihilate with an electron, making two oppositely-directed $0.511\,$MeV gamma rays.  Using the techniques of a PET scan, the location near (within a fraction of a millimeter with high probability) where the reacted thymine resided can be found. We will see later that the location for maximal production of F18 is typically a fraction of a millimeter behind a pristine Bragg peak location. (See end of Appendix \ref{sec-K-z}.)

The beta decay of the fluorine,
\begin{equation}
(\,^{18}\text{F}) \rightarrow (\,^{18}\text{O})^{-} +e^{+}+\nu _{e}\ ,
\label{F-beta}
\end{equation}
is the dominant (97\%) cause of the finite lifetime of $\,^{18}\text{F}$ ($\tau_{1/2}=109.77$\,min), with about 3\% occurrence of electron capture (also via a weak interaction).  (The parenthesis in the displayed reaction `equation' indicates the atom rather than just the nucleus.) 

Both the reaction (\ref{F-beta})
and the electron capture reaction  
\begin{equation}
(\,^{18}\text{F})^++(e^{-})\rightarrow (\,^{18}\text{O})+\nu_e\,
\label{F-EC}
\end{equation}
are transitions of $\,^{18}\text{F}[\,1^{+}]$ directly to the nuclear ground state of $\,^{18}\text{O}[\,0^{+}]$ (so there will be no subsequent gamma emission).
The bracketed notation shows the spin-parity of the nucleus. For the beta-decay reaction,
with the nuclear angular momentum $L=0$ in the final state, the positron-neutrino system will be in a triplet (negative parity) spin state, so the nuclear transition must be mediated by an odd parity operator, making this reaction an allowed Gamow-Teller transition.

Generally, a nuclear beta decay can be represented as a reaction:
\begin{equation}
\mathcal{N}^*(A,Z)\rightarrow \mathcal{N}(A,Z{\mp}1)+e^{\pm}+\nu^{\prime}_e \ ,
\end{equation}
where, for positron emission, $\nu^{\prime}_e = \nu_e$, i.e. a left-handed neutrino with lepton number $1$, while for electron emission,  $\nu^{\prime}_e = \overline{\nu}_e$, i.e. a right-handed anti-neutrino with lepton number $-1$. $\mathcal{N}^*(A,Z)$ is the `parent' nucleus with mass number $A$ (the number of protons plus the number of neutrons; used as
an isotope label) and nuclear charge $Z\left|e\right|$, while $\mathcal{N}$ is the daughter nucleus.
If the daughter nucleus is produced in an excited state, a 'prompt' gamma ray is also likely to be emitted from the daughter, although other channels for the fast nuclear decay of an excited nucleus are possible, such as $\alpha$-particle emission, proton or neutron emission, or even fission. (These daughter decays into nuclear particles are referred to as `beta-delayed' nuclear decays.) Daughters may also undergo another beta decay.

\begin{figure}[thpb]
\caption{Positron number per unit with energy (\citet[p.786]{levin1999})}
\center
\includegraphics[width=0.7\columnwidth]{ 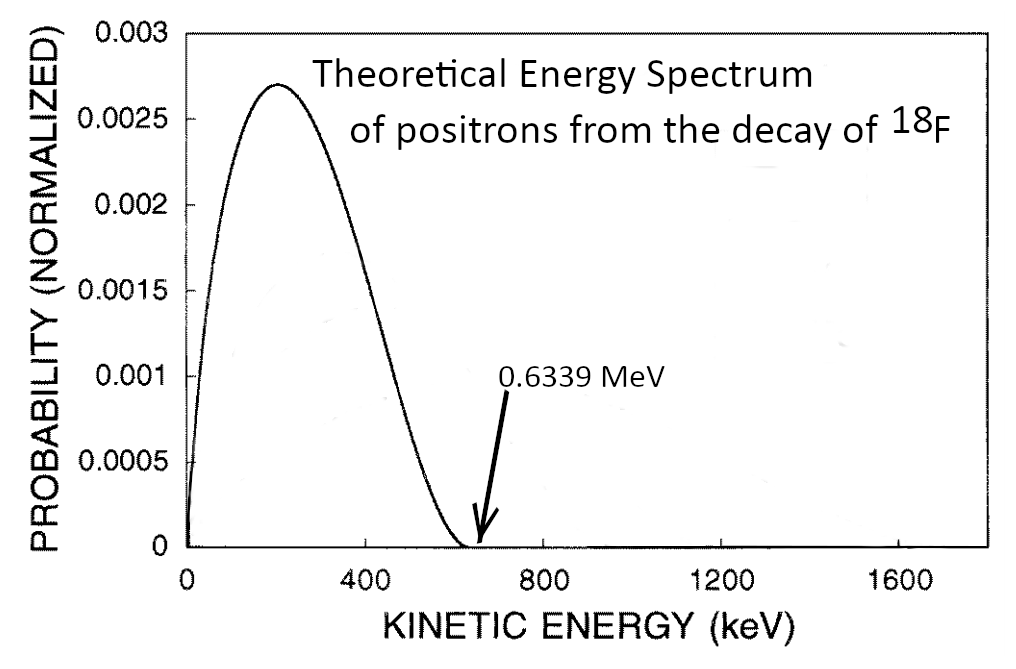}
\label{fig-1}
\end{figure}

In the case of F18 decay, the experimentally-measured maximum positron kinetic energy is $K_{max}=0.634$ MeV, with an average of about a third of this maximum. This kinetic energy
can be calculated using energy-momentum conservation, as shown in Appendix \ref{sec-fate}.  Including the Coulomb repulsion that the positron feels on exiting the daughter nucleus,  the number of positrons $N(E)dE$ produced in the weak decay in a given energy interval $dE$, when the
positron energies have energy centered on $E=K+m_ec^2$ is given by (\citet{wu}):
\begin{equation}
N\left( E\right) dE=gF(Z,E)pE\left( E_{\max }-E\right) ^{2}dE \ .
\end{equation}
Here, $g$ is a coupling constant, and $F\left( Z,E\right) $ is the Coulomb Fermi
function (coming from the magnitude squared of the overlap of the wave functions for the two leptons), with $E_{\max }=K_{max}+m_ec^2.$
Figure \ref{fig-1} shows the probability per unit energy,  $N(E)$, for the emission of a positron.

\begin{figure}[thpb]
\caption{Positron tracks (from \citet[p.792]{levin1999})}
\center
\includegraphics[width=0.6\columnwidth]{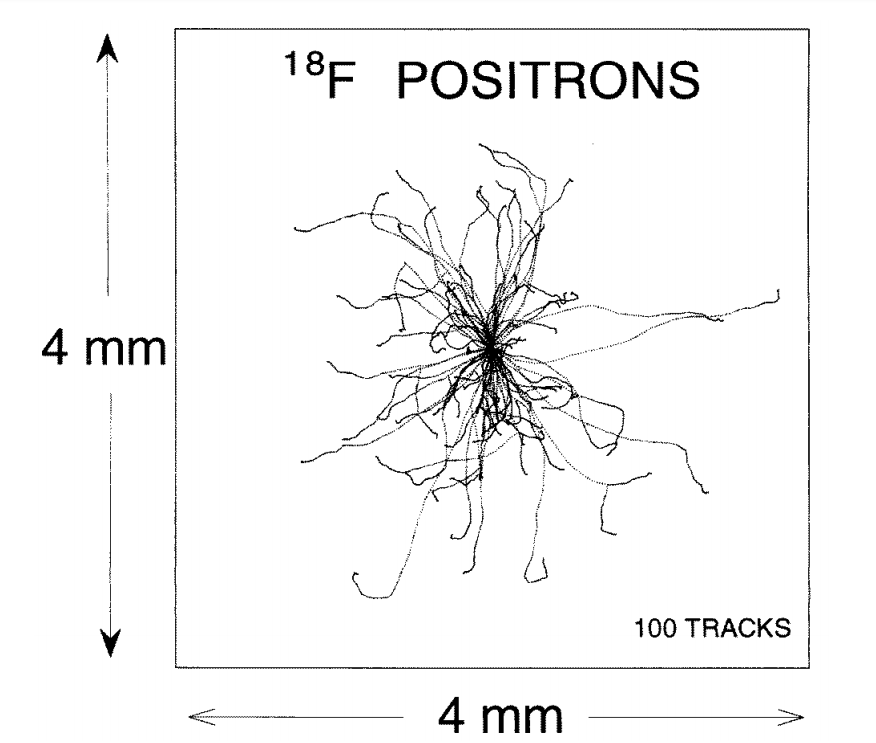}
\label{fig-2}
\end{figure}

\begin{figure}[thpb]
\caption{Positron spread around ``Line of Response'' (from \citet[p.793]{levin1999}).}
\includegraphics[width=1.0\columnwidth]{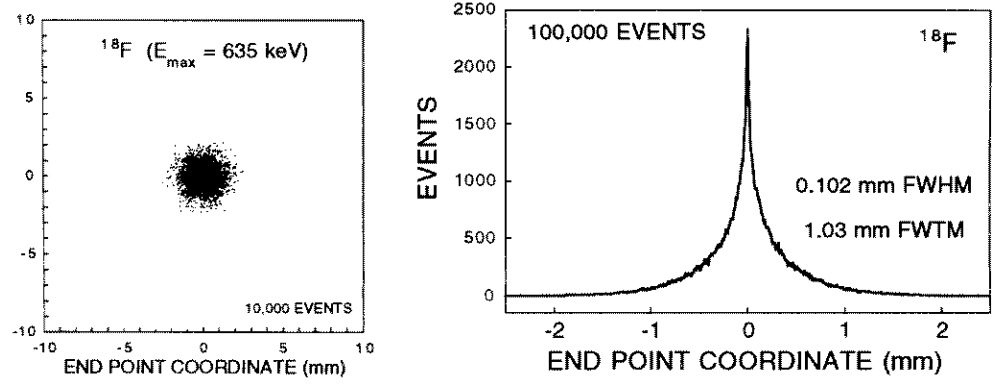}
\label{fig-2b}
\end{figure}

Figure \ref{fig-2} shows simulated positron tracks from a series of beta decays at one localized point.
At the end of each track, the positron combines with a local electron to make two
gamma rays (oppositely directed) of energy $m_{e}c^{2}=0.511\,$MeV
each. These gamma rays have an inverse attenuation coefficient of about $10\,$cm (\citet{kinahan2003x}) in water and
soft tissue, so, if produced in a body, they will likely exit and can be simultaneously detected in a PET scanner.

After the proton-beam induces  O18 (in a double-covalent bond with carbon \#2 in the thymine ring) to become F18 with the release of a neutron, the recoiling fluorine has sufficient energy to directly or indirectly break both backbone strands of the nearby DNA  (see Appendix \ref{app-recoil}), or, if it
remains in the nucleotide, causes the nitrogen \#3 in the thymine ring to form a double bond with carbon \#2, and drop the hydrogen at the nitrogen \#1.  That hydrogen formed one of two hydrogen bonds that kept the thymidine attached to adenocine. Each of the possible actions disrupt the DNA. With a F18 half-life of $109.77$ min, the electrons around the new fluorine nucleus have plenty of time to settle into stable orbits. Later,
the fluorine nucleus beta decays, with the positron quickly leaving the scene. Now the atomic electrons find themselves in orbit around the new oxygen nucleus, with one more electron than needed to make the atomic states of neutral oxygen, and with transient energies lower than those for the new stable oxygen orbits, since the number of protons at the core of the atom is one less. As the electrons settle up, they will have to release their excess energy to adjacent atoms, directly or by radiation (in the $eV$ range). From the complete ionization energies of fluorine and oxygen (106,434 keV and 84.078 keV) and by estimating the binding energy of the last electron in negatively charged oxygen (electron affinity of about 1.46 eV), there will be about 22 keV released by the new charged oxygen in becoming neutral.

\section{Nuclear reactions along the proton beam}

Protons and secondary neutrons in the PT beam can undergo nuclear interaction with the tissue nuclei via elastic, inelastic, and nonelastic collisions. 
For proton beams entering biological tissue, collisions with H1, O16, C12, and N14 will dominate, as these are the most abundant nuclides. 
At the highest beam energies ($\sim 250\,$MeV), a beam particle has time to interact with only one or two nuclei before reaching its range in the tissue. Because of Lorentz contraction, the beam particle acts as if the nucleus were flattened (perpendicular to the beam-particle's momentum), and the particles within as hardly moving and weakly held small-target dots.  Bremsstrahlung photon emission is negligible for therapeutic proton beams. At intermediate energies (tens of MeV), the beam protons and secondary neutrons can knock out protons, neutrons, and alpha particles, and multiple scattering within the nucleus can occur. Even nuclear fission can be induced.

The cross sections for proton-proton scattering and nuclear reactions produced by the proton beam are much smaller than
the electron-proton scattering cross sections, and even more so when the proton energy is much above the resonance region of the nuclear cross sections.
However, when the protons are slowed into the lower MeV range, i.e. in front of the Bragg peak, nuclear scattering and transmutations occur with higher probability.  
Near the end of the beam-protons range, the nuclear reactions largely produce `compound' nuclides, wherein the proton energy is shared among a number of nucleons in the nucleus before that nucleus decays. Excitation of collective motion of the nucleons is possible, producing `giant resonances' in the cross sections.


The proton fluence drops with depth in the tissue because of these reactions. In addition, proton-proton scattering
causes a spread of the beam.  
The radial spread has a standard deviation of about $2\,$\% of the beam range, causing a spread in dose delivery.  Beam intensities can produce over $1\,$Gy/min within the Bragg peak. Because protons at the far end of their range lose energy faster than at any other location, uncertainties in the range of the proton beam can cause a large uncertainty (estimated to be about $\pm 2.5\,$\% of the range \cite{paganetti2012range}) in the dose delivery at the far end of the Bragg peak.

An indication of the relative importance of nuclear secondaries is the fraction of the initial proton energy that
is apportioned to each kind of secondary. For inelastic nuclear interactions, using a Monte Carlo calculation, Seltzer \cite{seltzer1993} finds the following fractions for the reaction
$p+ \,_{O}^{16}O\rightarrow N^* +$secondaries:
\begin{center}
\renewcommand{\arraystretch}{1.4}
\begin{tabular}{CCCCCCCC}
\hline
$beam proton$ & &\multicolumn{3}{C}{$secondary particles$}&&&$$N^*\\
$energy$ &p\ \ &n\ \ &d\ \ &\,_1^3H&\,_2^3He&\,_2^4He&$recoils$\\
\hline
250\,$MeV$&0.66&\ \ 0.21\ \ &\ \ 0.005\ \ &\ \ 0.002\ \ &\ \ 0.001\ \ &0.019\ \ &\ \ 0.009\\
150\,$MeV$&0.57&\ \ 0.2\ \ &\ \ 0.106\ \ &\ \ 0.002\ \ &\ \ 0.002\ \ &0.029\ \ &\ \ 0.016\\
50\,$MeV$&0.36&\ \ 0.073\ \ &\ \ 0.051\ \ &\ \ 0.001\ \ &\ \ 0.003\ \ &0.098\ \ &\ \ 0.039\\
10\,$MeV$&0.17&\ \ 0.0\ \ &\ \ 0.0\ \ &\ \ 0.0\ \ &\ \ 0.0\ \ &0.11\ \ &\ \ 0.085\\
\hline
\end{tabular}
\end{center}
the remaining energy fraction going to elastic recoils and gamma rays. The emitted photons and the neutrons are
likely to travel a distance much larger than the transverse dimension of the proton beam.


Table \ref{tab-reactions} shows nonelastic nuclear reactions that can occur for O16 and O18
in tissue:
\begin{table}[!htbp]
\center
\caption{Important nonelastic reactions}
\label{tab-reactions}
\renewcommand{\arraystretch}{1.2}
\begin{tabular}{RCLCCLCR}
\hline\hline
\multicolumn{2}{L}{$Nuclear Reaction$}&&Q($MeV$)&$Decay\ of\ N$^*&$Half-life$&E_{EC}($MeV$)&E_e^{max}($MeV$)\\[2pt]
\hline
p+\,_{8}^{16}O&\rightarrow& \,_{9}^{17}F&+0.60027&\,_8^{17}O+\beta^+&64.49\,$sec$&2.7607&1.7387\\[2pt]
p+\,_{8}^{16}O&\rightarrow& n+\,_{9}^{16}F&-16.200&\,_{8}^{16}O+p&prompt&\multicolumn{2}{L}{(Q'=+0.536\,$MeV$)}\\[2pt]
p+\,_{8}^{16}O&\rightarrow& n+p+\,_{8}^{15}O&-15.664&\,_{7}^{15}N+\beta^+&2.04\,$min$&2.7539&1.7320\\[2pt]
p+\,_{8}^{16}O&\rightarrow& p+p+\,_{7}^{15}N&-12.127&stable&&\\[2pt]
p+\,_{8}^{16}O&\rightarrow& n+n+p+\,_{8}^{14}O&-2.887&\,_7^{14}N+\beta^+&70.61\,$sec$&5.1430&4.1210\\[2pt]
p+\,_{8}^{16}O&\rightarrow& \alpha+\,_{7}^{13}N&-5.218&\,_6^{13}C+\beta^+&9.965\,$min$&2.2204&1.1984\\[2pt]
p+\,_{8}^{16}O&\rightarrow& p+\alpha+\,_{6}^{12}C&-7.162&stable&&\\[2pt]
p+\,_{8}^{16}O&\rightarrow& d+\,_{8}^{15}O&-13.439&stable&&\\[2pt]
\\
p+\,_{8}^{18}O&\rightarrow& \,_{9}^{19}F&+7.9936&stable\\[2pt]
p+\,_{8}^{18}O&\rightarrow& n+\,_{9}^{18}F&-2.438&\,_{8}^{18}O+\beta^+&109.77\,$min$&1.6555&0.6335\\[2pt]
p+\,_{8}^{18}O&\rightarrow& n+p+\,_{8}^{17}O&-8.045&stable&&\\[2pt]
p+\,_{8}^{18}O&\rightarrow& p+p+\,_{7}^{17}N&-15.942&\,_8^{17}O+\beta^-&4.173\,$sec$&&8.6800\\[2pt]
p+\,_{8}^{18}O&\rightarrow& n+n+p+\,_{8}^{16}O&-12.188&stable&&\\[2pt]
p+\,_{8}^{18}O&\rightarrow& \alpha+\,_{7}^{15}N&+3.980&stable&\\[2pt]
p+\,_{8}^{18}O&\rightarrow& d+\,_{8}^{17}O&-5.821&stable&&\\[2pt]
\\
n+\,_{8}^{16}O&\rightarrow& \,_{8}^{17}O&+4.143&stable&\\[2pt]
n+\,_{8}^{16}O&\rightarrow& p+\,_{7}^{16}N&-9.639&\,_8^{16}O+\beta^-&7.13\,$sec$&&10.4200\\[2pt]
n+\,_{8}^{16}O&\rightarrow& p+n+\,_{7}^{15}N&-12.127&stable&&\\[2pt]
n+\,_{8}^{16}O&\rightarrow& n+n+\,_{8}^{15}O&-15.664&\,_{7}^{15}N+\beta^+&2.04\,$min$&2.7539&1.7319\\[2pt]
n+\,_{8}^{16}O&\rightarrow& p+p+\,_{6}^{15}C&-21.117&\,_{7}^{15}N+\beta^-&2.45\,$sec$&&9.7717\\[2pt]
\\
n+\,_{8}^{18}O&\rightarrow& \,_{8}^{19}O&+3.956&\,_9^{19}N+\beta^-&26.91\,$sec$&&4.8210\\[2pt]
n+\,_{8}^{18}O&\rightarrow& p+\,_{7}^{18}N&-13.114&\,_8^{18}O+\beta^-&624\,$msec$&4.4334&3.4114\\[2pt]
n+\,_{8}^{18}O&\rightarrow& p+n+\,_{7}^{17}N&-15.942&\,_8^{17}O+\beta^-&4.173\,$sec$&&8.6808\\[2pt]
n+\,_{8}^{18}O&\rightarrow& n+n+\,_{8}^{17}O&-8.045&stable&&\\[2pt]
n+\,_{8}^{18}O&\rightarrow& p+p+\,_{6}^{17}C&-28.321&\,_8^{17}O+2\beta^-&4.2\,$sec$&&21.8460\\[2pt]
\hline
\end{tabular}
\end{table}
The $Q$ values in Table \ref{tab-reactions}, i.e. the energy released from mass, are given using atomic masses, not fully ionized isotopes. (See Appendix \ref{app-Q}. The $Q$ values come from \citet{wang2017} as reported by \cite{NNDC2016}.) Any of these reactions
may involve the release of a gamma ray, but gammas are unlikely to ionize locally.

The production of tritium and helium-3 occurs at a much lower rate, as the nucleons in these
products do not form tight clusters in heavier nuclei, and so these clusters are not as likely to be knocked out
as a particle. Slow neutrons with tissue hydrogen will also make deuterium, but this will occur most often outside the target tissue. As the table shows, many of the daughter nuclei ($\mathcal{N}^*$) after the `strong' nuclear interaction are radioactive.

{\bf Note the striking difference} in the reaction $Q$ values between the list of O16 vs O18 reactions
shown in Table \ref{tab-reactions}. Corresponding reactions of the proton with O18 show far less energy loss in the release of the reaction products. Two of the O18 reactions are strongly exoergic and these are among the largest in cross section, while for the proton-O16 reactions, only the direct capture is exoergic, but with only 600\,keV released compared to almost $8\,$MeV for O18. Overall, far more energy will be available for kinetic energy of ionizing particles in the case of $p+O18$ than for $p+O16$. The root cause is the tighter binding of O16, being a doubly `magic' nucleus, with the lowest lying nuclear states for neutrons and protons being fully occupied, and with an energy gap to the next level. 

The reactions that produce radionuclides that are positron emitters may add $0.511\,$MeV
gamma emissions during a PET scan, indistinguishable from the F18 decay gammas. (See Appendix \ref{app-exam}.)
Beside the ones listed above, there are other nuclear reactions in tissue that contribute to positron production. Important ones are shown in Table \ref{tab-react2}.
\begin{table}[!htbp]
\center
\caption{Other contributing positron-producing reactions}
\label{tab-react2} 
\renewcommand{\arraystretch}{1.4}
\begin{tabular}{RCLCC}
\hline\hline
\multicolumn{2}{L}{$Nuclear Reaction$}&&$Decay of$N^*&$\ $N^*$ Half-life$\\[2pt]
\hline 
p+\,_{6}^{16}O&\rightarrow&d+\,_{6}^{15}O&\,_{7}^{15}N+\beta^+&2.04\,$min$ \\[2pt]
p+\,_{6}^{16}O&\rightarrow&p+n+\,_{6}^{15}O&\,_{7}^{15}N+\beta^+&2.04\,$min$ \\[2pt]
p+\,_{6}^{16}O&\rightarrow&d+\alpha+\,_{6}^{11}C&\,_{5}^{11}B+\beta^+&20.4\,$min$ \\[2pt]
p+\,_{6}^{16}O&\rightarrow&p+n+\alpha+\,_{6}^{11}C&\,_{5}^{11}B+\beta^+&20.4\,$min$ \\[2pt]
p+\,_{6}^{12}C&\rightarrow& n+\,_7^{12}N&3\alpha+\beta^+&11.0\,$sec$\\[2pt]
p+\,_{6}^{12}C&\rightarrow& d+\,_{6}^{11}C&\,_{5}^{11}B+\beta^+&20.4\,$min$\\[2pt]
p+\,_{6}^{12}C&\rightarrow& p+n+\,_{6}^{11}C&\,_{5}^{11}B+\beta^+&20.4\,$min$\\[12pt]
n+\,_{6}^{12}C&\rightarrow& n+n+\,_{6}^{11}C&\,_{5}^{11}B+\beta^+&20.4\,$min$ \\[8pt]
p+\,_{7}^{14}N&\rightarrow& d+\,_{7}^{13}N&\,_{6}^{13}C+\beta^+&9.965\,$min$ \\[2pt]
p+\,_{7}^{14}N&\rightarrow& p+n+\,_{7}^{13}N&\,_{6}^{13}C+\beta^+&9.965\,$min$ \\[2pt]
p+\,_{7}^{14}N&\rightarrow& \alpha+\,_{6}^{11}C&\,_{5}^{11}B+\beta^+&20.39\,$min$ \\[8pt]
p+\,_{11}^{23}Na&\rightarrow&d+\,_{11}^{22}Na&\,_{10}^{22}Ne+\beta^+&2.602\,$yrs$ \\[8pt]
p+\,_{11}^{23}Na&\rightarrow&p+n+\,_{11}^{22}Na&\,_{10}^{22}Ne+\beta^+&2.602\,$yrs$ \\[8pt]
n+\,_{11}^{23}Na&\rightarrow&n+n+\,_{11}^{22}Na&\,_{10}^{22}Ne+\beta^+&2.602\,$yrs$ \\[8pt]
p+\,_{9}^{19}F&\rightarrow& d+\,_{9}^{18}F&\,_8^{18}O+\beta^+&1.8295\,$hrs$ \\[6pt]
p+\,_{9}^{19}F&\rightarrow& p+n+\,_{9}^{18}F&\,_8^{18}O+\beta^+&1.8295\,$hrs$ \\[6pt]
n+\,_{9}^{19}F&\rightarrow& n+n+\,_{9}^{18}F&\,_8^{18}O+\beta^+&1.8295\,$hrs$ \\[2pt]
\hline
\end{tabular}
\end{table}

\begin{figure}[thpb]
\caption{Cross section for F19$(p,np)F18$ (plotted by EXFOR \cite{zerkin} from \citet{yule1960,marquez1952}) }
\label{fig-F19F18}
\center
\includegraphics[width=0.7\columnwidth]{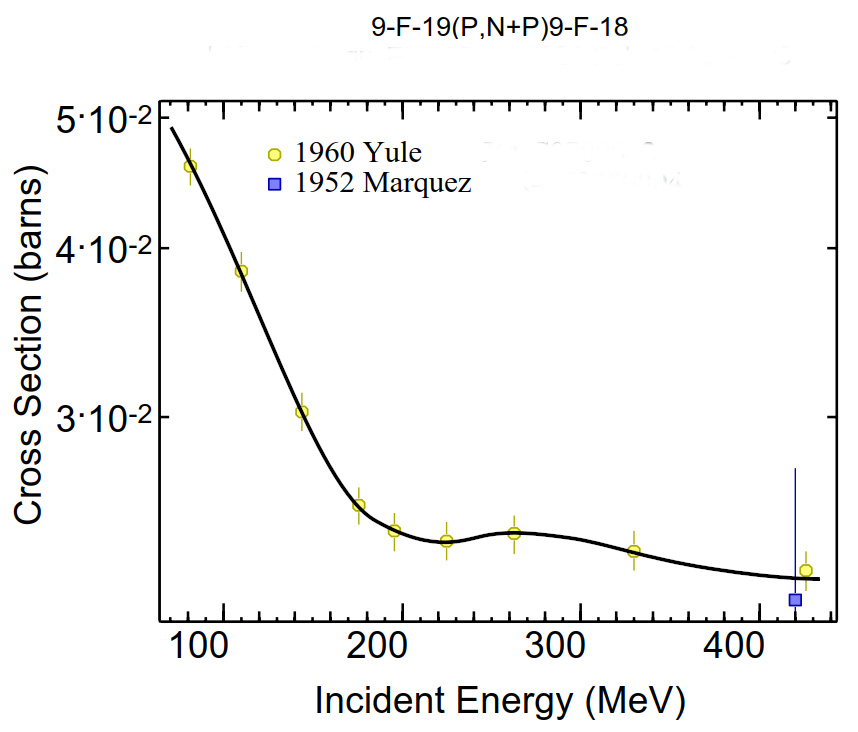}
\label{fig-F19-F18}
\end{figure}

When a tumor is doped with O18, the reaction O18$(p,n)$F18 will occur in PT therapy. 
The produced F18 half-life is $109.7\,$min. If a PET scan occurs an hour or so past 
the PT therapy, $\beta^{+}$-induced gamma emissions from the radionuclides in the 
above table will have died down except for the F18 produced from fluorine in the tissue, 
and from Na22. 
The concentration of natural sodium Na23 in tissue is low ($0.037\,$\% by atoms 
or $3.7\times 10^{19}\,$ ions/cm$^3$), and Na22 has a long half-life ($2.6\,$years), 
indicating that Na23 activity will be low compared to F18.  The production of F18 from the fluorine in tissue would directly
interfere with PET after PT with doped DNA.  However, the concentration of
fluorine in tissue is about $1.2\times 10^{18}\,$ atoms/cm$^3$. As shown 
in Table \ref{tab-thymine}, the number density of F18 after a full PT is about
$8$ to $24 \times 10^{21}$ per cubic centimeter. Moreover, the cross section 
for F19$(p,d)$F18 is about $100$ times lower than O18$(p,n)$F18 (see Fig.\,\ref{fig-F19F18}).
The reaction F19$(n,nn)$F18 contributes even less, since the neutron flux is 
lower than the proton beam flux. Thus, the PET signal from fully exposed thymine-18 
should be far larger than that from the F18 produced by proton-converted naturally-occurring F19. 

\section{Cross section for O18\,$\MakeLowercase{(n,p)}$\,F18}

\begin{figure}[thpb]
\caption{Total cross section for O$(p,X)$ from \citet{ulmer2010}}
\center
\includegraphics[width=0.8\columnwidth]{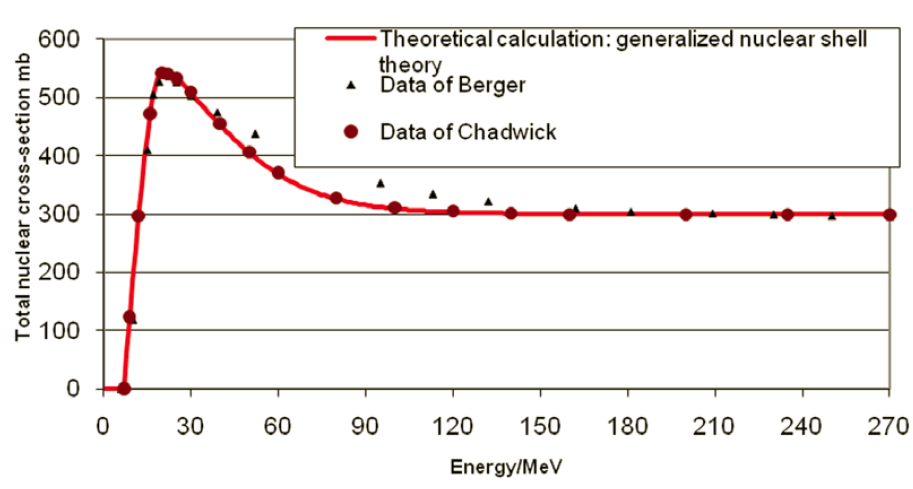}
\label{fig-O16-total}
\end{figure}

The cross section for  O18$(p,n)$F18 is shown in the two graphs below.
\begin{figure}[thpb]
\caption{Cross section for O18$(p,n)$F18 2 to 20 MeV (From Cyclotron Produced Radionuclides: Principles and Practice, (2008))}
\center
\includegraphics[width=1\columnwidth]{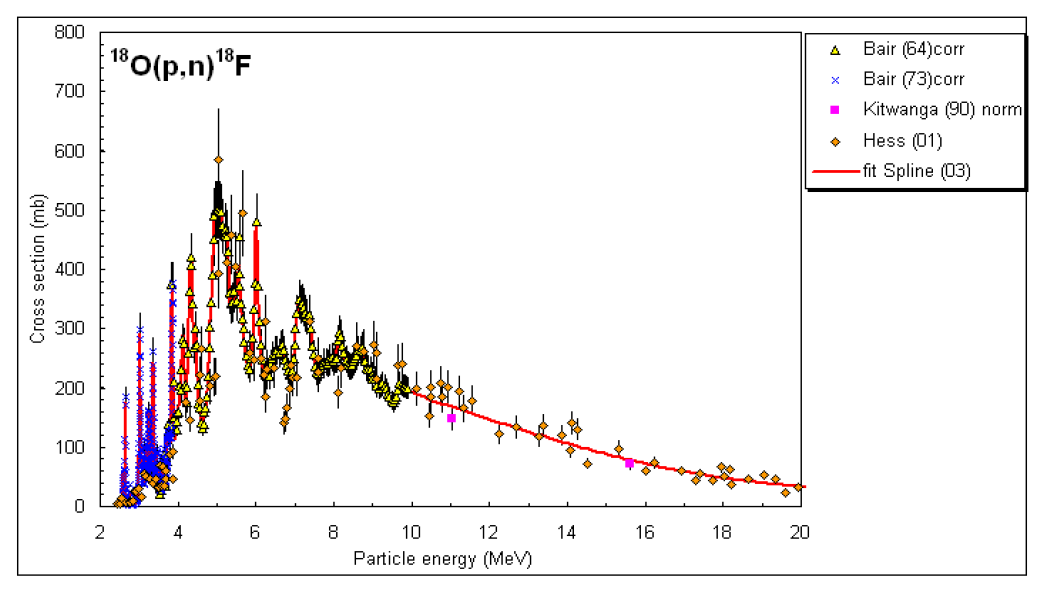}
\label{fig-3}
\end{figure}
\begin{figure}[!ht]
\caption{Cross section for O18$(p,n)$F18 0 to 30 MeV, from \citet[p361]{hess2001}}
\center
\includegraphics[width=0.8\columnwidth]{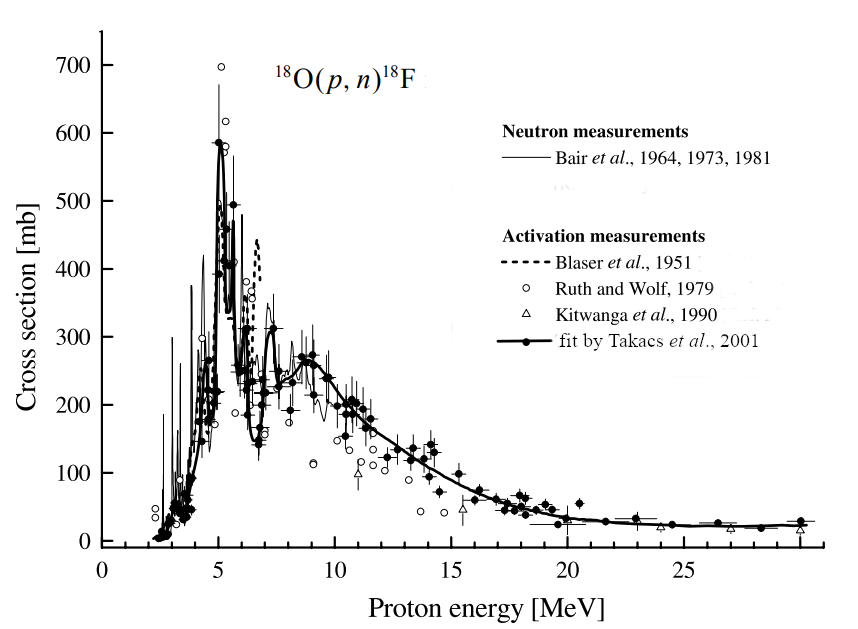}
\label{fig-O18F18-cross}
\end{figure}
For comparison, Fig.\,\ref{fig-O16-total} shows the total cross section for protons
impinging on oxygen-16.
Among all the reactions shown in Table \ref{tab-reactions}, the cross section for
O18$(p,n)$F19 is large due to a number of resonances in the proton energy
region of $4$ to $12\,$MeV.

\section{Estimating the increase in dose with O18 substituted for 16O in DNA thymine}
\label{sec-dose}

To estimate the increase in dose due to inelastic nuclear reactions with O18 instead of O16, one can first calculate the LET due to the reaction O18$(p,X)$
and compare to O16$(p,X)$.
This is an integration over energy for each $p+\mathcal{N}$ reaction cross section as a function of energy times an average energy given to charged-particles in that reaction times the number density of the reactant times the flux rate of the beam protons at the reaction location.

There are several important factors that must be considered:
\begin{itemize}
\item The energy and flux of protons in the beam at the location of the reaction;
\item The cross section for the reactions O18$(p,X)$;
\item The density of O18 in the exposed tissue;
\item The energies of the released charged particles.
\end{itemize}
The consequent LET per unit area per unit time per length along the beam has the general form:
\begin{equation}
\label{eq-dK}
\frac{dK}{dAdt\,dz}=-J\sum_i \rho_i \sum_j\int_{\Omega_j}{\frac{d\sigma_{ij}}{d\Omega_j} \epsilon_{ij}(K,\Omega_j)d\Omega_j} \ ,
\end{equation}
where $\sigma_{ij}(K)$ is the cross section for the reaction with reactant index $i$ and exiting particle $j$ at beam energy $K$; 
$\epsilon_{ij}(K,\Omega_j)$ is the kinetic energy of a released charged particle labeled by $j$ and exiting into solid angle $\Omega_j$. The quantity
$\rho_i$ is the number density of reactants in the tissue 
and $J$ is the number current density of protons in the beam, both at the location where the beam has energy $K$. The expression Eq.\,(\ref{eq-dK}) can be handled in a computer calculation, drawing on a database of reactions and cross sections, such as GEANT4 \cite{allison2006geant4}.

\section{Summary}

The substitution of O18 for O16 in tissue thymine causes a greater PT dose to be delivered to cancer cells, a conclusion one can draw from the fact that the important inelastic nuclear reactions with O16 are far more endoergic than the corresponding reactions with O18. The delayed beta decay of F18 adds more to the
dose delivered, and makes it possible to know, within about a millimeter, where the delivery occurred, using PET.
Given these facts, a full calculation of the numerical $RBE$ with O18 is justified.


\appendix

\section{Some Relevant Data}

Some selected physical masses are given in Table \ref{tab-datam} in daltons and energy units of MeV (with time units to make $c=1$). The column labeled "Complete
Ionization" comes from the CRC ``Ionization energies of atoms and atomic ions'' data \cite{crc}, converted to eV. The values for the bare isotopes of O16, O18 and F18 are calculated from the complete ionization energy and the atomic mass-energy using
\begin{equation}
m_{bare}=m_{atom}-Z\,m_e+I \ ,
\end{equation}
where $Z$ is the number of protons in the nucleus and $I$ is the complete ionization energy of the atom (energy to remove all the atom's electrons). 

The daughter nucleus of oxygen-18 is created with an extra atomic electron close-by.  As such, that electron
will have an attractive Coulomb force acting on it. The extra electron will likely take energy from other atomic electrons as they all shuffle to new stable oxygen-18 orbitals, higher in energy than the fluorine-18 orbitals they initially occupied.

\vspace{0.1in}

\begin{table}[ht]
\vspace{-.1in}
\caption{Selected Physical Masses (\citet{NIST-isotopes})}
\vspace{0.1in}
\label{tab-datam}
\renewcommand{\arraystretch}{1.4}
\begin{tabular}{llll}
\hline
\hline
& Mass & $Mc^2$ & \ \ Complete\\
& (Daltons) & (MeV) & \ \ Ionization (eV)\\
\hline
$m_p$ & 1.072764667 & 938.272088 &\\
$m_e$ & 0.000548579909 & 0.51099895 & \\
\hline
atom & & & \\
\hline
 (O16) & 15.99491461957 &14899.1686 & \ \ \ \ \ \ 871.4101 \\
 (O18) & 17.99915961286 & 16766.111 & \ \ \ \ \ \ 871.4101 \\
 (F18) & 18.00093733 & 16767.767 & \ \ \ \ \,1103.1176 \\
\hline
bare nucleus  & & &\\
\hline
  O16 & & 14895.0815 &\\
  O18 & & 16672.0239 & \\
  F18 & & 16763.1691 &\\
\hline
1 Dalton & \,\,=\,931.49410242(28) & MeV \\
\hline
\hline
\end{tabular}
\end{table}

In addition, the following facts are relevant: 

\vspace{0.1in}

The isotope O18 occurs naturally with a number ratio O18/O16\,$=2.05(14)\times 10^{-3}$ (\cite{hampel1968})
that varies from $(1.88$ to $2.22)\times 10^{-3}$. The greater amount is in
seawater due to slower evaporation of water containing O18 compared to 
O16. Thus, the percent of  O18 in a patient depends on the diet of that patient. A vegetable diet would favor an atmospheric abundance in tissue about $1.9\times 10^{-3}$ for  O18/O16, while a dominance of sea food will make $2.2\times 10^{-3}$ for  O18/O16. The density of O16 in soft tissue is dominated by that in the tissue water.  Specific element mass densities in tissue are given in Table \ref{tab-m-dense}.

\vspace{0.1in}

The F18 nucleus is a spin-parity $J^{\pi }=1^{+}$ system; the O18 nucleus
is a spin-parity $J^{\pi }=0^{+}$ system. These mean that the weak $\beta^+$ decay proceeds via an allowed Gamov-Teller transition, with the transition probability proportional to the magnitude square of the nuclear transition matrix element, a sum over all the nucleons, giving 
\[\left|{\textstyle \sum_{\kappa}\left<\Psi_{O18}\right|\vec{\sigma}_N(\kappa})\tau^{-}_N(\kappa)\left|\Psi_{F18}\right>
\cdot\chi^{\dagger}_{\nu}\vec{\sigma}_L(\kappa)\chi_{e}\right|^2\ ,\] 
with the nucleon spin operator $\vec{\sigma}_N$, lepton spin operator $\vec{\sigma}_L$ and the nucleon isospin lowering operator $\tau^{-}_N$. The symbol $\kappa$ labels individual nucleons in the nucleus. The $\chi$ are lepton spin states. After summing over the unobserved
spin states of the leptons, the transition probability is proportional to
\[\left|\left<\Psi_{O18}\right|{\textstyle \sum_{\kappa}}\vec{\sigma}_N(\kappa)\tau_{N}^{-}(\kappa)\left|\Psi_{F18}\right>\right|^2\ ,\]
where a vector dot-product is implied between the two matrix-element factors.  We see that
this beta decay will change a proton into a neutron, and, given the spin-parity
of O18 and F18, will flip its spin.

\vspace{0.1in}

The lifetime of F18 is $109.771(20)\,$min, $97$\% by positron emission, 
$3$\% by electron capture. Other positron emitters that can be used in PET scans are shown in Table \ref{beta-emit}. Kmax
and Kmean refer to the positron kinetic energy maximum and mean.  Rmax and Rmean are ranges of the positron. Of those shown, F18 has the best combination of half-life and positron range for PET scan application.

\begin{table}[h]
\caption{Some properties of beta-emitters used in PET (from \citet{conti2016})}
\renewcommand{\arraystretch}{1.4}
\begin{tabular}{lcccccc}
\hline
\hline
Isotope & Half-life & Branching ($\beta^+$) & Kmax (MeV) & Kmean (MeV) & Rmax (mm) & R mean (mm)\\
\hline
$\,^{11}$C&20.4 min&99.8 \%&0.960&0.386&4.2&1.2\\
$\,^{13}$N&10.0 min&99.8 \%&1.199&0.492&5.5&1.8\\
$\,^{15}$O&2 min&99.9 \%&1.732&0.735&8.4&3.0\\
$\,^{18}$F&110 min&96.9 \%&0.634&0.250&2.4&0.6\\
$\,^{64}$Cu&12.7 h&17.5 \%&0.653&0.278&2.5&0.7\\
$\,^{89}$Zr&78.4 h&22.7 \%&0.902&0.396&3.8&1.3\\
\hline
\end{tabular}
\label{beta-emit}
\end{table}

\vspace{0.1in}

Table \ref{beam-reactions} shows the most important positron emitters created by
a proton beam in tissue.

\begin{table}[!htbp]
\caption{Positron emitters created by proton beams in tissue (\citet{studenski2010})}
\renewcommand{\arraystretch}{1.4}
\begin{tabular}{lccc}
\hline
\hline
Reaction & Threshold energy & Half-life &Positron energy\\
 & MeV & min & MeV\\
\hline
$\,^{16}O$(p,pn)$^{15}$O & 16.79 & 2.037 & 1.72\\
$\,^{16}O$(p,$\alpha$)$\,^{13}$N & 5.66 & 9.965 & 1.19\\
$\,^{14}$N(p,pn)$^{13}$N & 11.44 & 9.965 & 1.19\\
$\,^{12}$C(p,pn)$^{11}$C & 20.61 & 20.390 & 0.96\\
$\,^{14}$N(p,$\alpha$)$\,^{11}$C & 3.22 & 20.390 & 0.96\\
$\,^{16}$O(p,$\alpha$pn)$^{11}$C & 59.64 & 20.390 & 0.96\\
\hline
\end{tabular}
\label{beam-reactions}
\end{table}

\vspace{0.1in}


\begin{table}[!htbp]
  \centering
  \caption{Percent by mass of principal elements in tissue and ionization energy (\citet[p.5]{hunemohr2014})}
\renewcommand{\arraystretch}{0.96}
    \begin{tabular}{lcccccccr}
    Material & \multicolumn{1}{l}{$\rho$ [g/cm$^3$]} & \multicolumn{1}{l}{\ \ H\ \ \ $\cdot$} & \multicolumn{1}{l}{\ \ C\ \ \ $\cdot$} & \multicolumn{1}{l}{\ N\ \ \ $\cdot$} & \multicolumn{1}{l}{\ O\ \ \ $\cdot$} & \multicolumn{1}{l}{\ Ca} & \multicolumn{1}{l}{\ \ P\ \ \ $\cdot$} & \multicolumn{1}{l}{I\ [eV]} \\
\hline
    \rule[-0.in]{0pt}{0.15in}Lung deflated & 0.26  & 10.4  & 10.6  & 3.1   & 75.7  & 0     & 0.2   & 74.54 \\
    Yellow marrow & 0.98  & 11.5  & 64.6  & 0.7   & 23.2  & 0     & 0     & 63.72 \\
    Mammary gland1 & 0.99  & 10.9  & 50.8  & 2.3   & 35.9  & 0     & 0.1   & 66.73 \\
    Mammary gland2 & 1.02  & 10.6  & 33.3  & 3     & 52.9  & 0     & 0.1   & 70.05 \\
    Mammary gland3 & 1.06  & 10.2  & 15.9  & 3.7   & 70.1  & 0     & 0.1   & 73.78 \\
    Red marrow & 1.03  & 10.6  & 41.7  & 3.4   & 44.2  & 0     & 0.1   & 68.7 \\
    Brain Cerebrospinal fluid & 1.01  & 11.2  & 0     & 0     & 88.8  & 0     & 0     & 75.3 \\
    Adrenal gland & 1.03  & 10.7  & 28.5  & 2.6   & 58.1  & 0     & 0.1   & 70.92 \\
    Smallintestine wall & 1.03  & 10.7  & 11.6  & 2.2   & 75.5  & 0     & 0.1   & 73.98 \\
    Urine & 1.02  & 11.1  & 0.5   & 1     & 87.2  & 0     & 0.1   & 75.21 \\
    Gall bladder bile & 1.03  & 10.9  & 6.1   & 0.1   & 82.9  & 0     & 0     & 74.81 \\
    Lymph & 1.03  & 10.9  & 4.1   & 1.1   & 83.9  & 0     & 0     & 75 \\
    Pancreas & 1.04  & 10.7  & 17    & 2.2   & 69.9  & 0     & 0.2   & 73 \\
    Brain white matter & 1.04  & 10.7  & 19.6  & 2.5   & 66.8  & 0     & 0.4   & 72.5 \\
    Prostate & 1.04  & 10.6  & 9     & 2.5   & 77.9  & 0     & 0.1   & 74.58 \\
    Testis & 1.04  & 10.7  & 10    & 2     & 77.2  & 0     & 0.1   & 74.23 \\
    Brain gray matter & 1.04  & 10.8  & 9.6   & 1.8   & 77.5  & 0     & 0.3   & 74.17 \\
    Muscle skeletal1 & 1.05  & 10.2  & 17.3  & 3.6   & 68.7  & 0     & 0.2   & 73.69 \\
    Muscle skeletal2 & 1.05  & 10.3  & 14.4  & 3.4   & 71.6  & 0     & 0.2   & 74.03 \\
    Muscle skeletal3 & 1.05  & 10.3  & 11.3  & 3     & 75.2  & 0     & 0.2   & 74.66 \\  
    Heart1 & 1.05  & 10.4  & 17.6  & 3.1   & 68.6  & 0     & 0.2   & 73.32 \\
    Heart2 & 1.05  & 10.5  & 14    & 2.9   & 72.4  & 0     & 0.2   & 73.8 \\
    Heart3 & 1.05  & 10.5  & 10.4  & 2.7   & 76.2  & 0     & 0.2   & 74.49 \\
    Heart blood filled & 1.06  & 10.4  & 12.2  & 3.2   & 74.1  & 0     & 0.1   & 74.22 \\
    Blood whole & 1.06  & 10.3  & 11.1  & 3.3   & 75.2  & 0     & 0.1   & 74.61 \\
    Kidney1 & 1.05  & 10.3  & 16.1  & 3.4   & 69.9  & 0.1   & 0.2   & 73.79 \\
    Kidney2 & 1.05  & 10.4  & 13.3  & 3     & 73    & 0.1   & 0.2   & 74.16 \\
    Kidney3 & 1.05  & 10.5  & 10.7  & 2.7   & 75.8  & 0.1   & 0.2   & 74.48 \\
    Stomach & 1.05  & 10.5  & 14    & 2.9   & 72.5  & 0     & 0.1   & 73.81 \\
    Thyroid & 1.05  & 10.5  & 12    & 2.4   & 75    & 0     & 0.1   & 74.24 \\
    Liver1 & 1.05  & 10.4  & 15.8  & 2.7   & 70.8  & 0     & 0.3   & 73.72 \\
    Liver2 & 1.06  & 10.3  & 14    & 3     & 72.3  & 0     & 0.3   & 74.18 \\
    Liver3 & 1.07  & 10.2  & 12.7  & 3.3   & 73.4  & 0     & 0.3   & 74.58 \\
    Aorta & 1.05  & 10    & 14.8  & 4.2   & 70.2  & 0.4   & 0.4   & 74.78 \\
    Ovary & 1.05  & 10.6  & 9.4   & 2.4   & 77.4  & 0     & 0.2   & 74.52 \\
    Eye lens & 1.07  & 9.6   & 19.6  & 5.7   & 64.9  & 0     & 0.1   & 73.97 \\
    Spleen & 1.06  & 10.4  & 11.4  & 3.2   & 74.7  & 0     & 0.3   & 74.47 \\
    Trachea & 1.06  & 10.2  & 14    & 3.3   & 72    & 0     & 0.4   & 74.38 \\
    Skin1 & 1.09  & 10.1  & 25.2  & 4.6   & 59.9  & 0     & 0.1   & 72.25 \\
    Skin2 & 1.09  & 10.1  & 20.6  & 4.2   & 65    & 0     & 0.1   & 73.17 \\
    Skin3 & 1.09  & 10.2  & 15.9  & 3.7   & 70.1  & 0     & 0.1   & 73.89 \\
    Connective tissue & 1.12  & 9.5   & 21    & 6.3   & 63.1  & 0     & 0     & 73.79 \\
    Cartilage & 1.1   & 9.8   & 10.1  & 2.2   & 75.7  & 0     & 2.2   & 76.96 \\
    Sternum & 1.25  & 7.8   & 31.8  & 3.7   & 44.1  & 8.6   & 4     & 81.97 \\
    Sacrummale & 1.29  & 7.4   & 30.4  & 3.7   & 44.1  & 9.9   & 4.5   & 84.19 \\
    Femur conical trochanter & 1.36  & 6.9   & 36.7  & 2.7   & 34.8  & 12.9  & 5.9   & 86.69 \\
    Sacrum female & 1.39  & 6.6   & 27.3  & 3.8   & 43.8  & 12.6  & 5.8   & 89.11 \\
    Humerus whole specimen & 1.39  & 6.7   & 35.3  & 2.8   & 35.3  & 13.6  & 6.2   & 88.06 \\
    Ribs 2nd to 6th & 1.41  & 6.4   & 26.5  & 3.9   & 43.9  & 13.2  & 6     & 90.28 \\
    Vert colC4 excl cartilage & 1.42  & 6.3   & 26.3  & 3.9   & 43.9  & 13.4  & 6.1   & 90.78 \\
    Femur total bone & 1.42  & 6.3   & 33.4  & 2.9   & 36.3  & 14.4  & 6.6   & 90.24 \\
    Femur whole specism & 1.43  & 6.3   & 33.2  & 2.9   & 36.4  & 14.5  & 6.6   & 90.34 \\
    Innominate female & 1.46  & 6     & 25.2  & 3.9   & 43.8  & 14.4  & 6.6   & 92.76 \\
    Humerus total bone & 1.46  & 6     & 31.5  & 3.1   & 37    & 15.3  & 7     & 92.23 \\
    Clavicle scapula & 1.46  & 6     & 31.4  & 3.1   & 37.1  & 15.3  & 7     & 92.26 \\
    Humerus cylindrical shaft & 1.49  & 5.8   & 30.3  & 3.2   & 37.6  & 15.9  & 7.2   & 93.56 \\
    Ribs10th & 1.52  & 5.6   & 23.7  & 4     & 43.7  & 15.7  & 7.3   & 95.42 \\
    Cranium & 1.61  & 5     & 21.3  & 4     & 43.8  & 17.7  & 8.1   & 99.69 \\
    Mandible & 1.68  & 4.6   & 20    & 4.1   & 43.8  & 18.8  & 8.7   & 102.35 \\
    Femur cylindrical shaft & 1.75  & 4.2   & 20.5  & 3.8   & 41.8  & 20.3  & 9.4   & 105.13 \\
    Cortical bone & 1.92  & 3.4   & 15.6  & 4.2   & 43.8  & 22.6  & 10.4  & 111.63 \\
    \end{tabular}%
  \label{tab-m-dense}%
\end{table}

\begin{table}
\centering
\caption{Thymine-18 in doped tissue (\cite{piovesan2019length},\cite{panajotovic2006effective})}
\label{tab-thymine}
\renewcommand{\arraystretch}{1.2}
\label{tab-O18-data}
 \begin{tabular}{lll}
\hline
 Mass of DNA && male 6.41\,pgram, female 6.51\,pgram\\
 Todal DNA length && male: 205.0\,cm, female 208.23\,cm\\
 Total base pairs && male: 6.27$\times 10^{12}$, female 6.37$\times 10^{12}$\\
A-T base pairs in DNA && $59.2$\%\\
Thymines in one cell && $0.592\times 6.3\times 10^{12}=3.8\times 10^{12}$\\
Cell diameter/nucleus diameter&&$\sim\, 2.3$\\
Tissue cell number density &&  $\rho_{cell} =\,$1 to 3 billion cells per cm$^3$\\
Density of thymine in tissue &&  $\rho_{Thy} =\,$4 to 12 $\times 10^{21}$ per cm$^3$\\
Density of O18 in dopped tissue&& $\rho_{O18} =\,$4 to 12 $\times 10^{21}$ per cm$^3$\\
Cross section each O18 && $\sigma_{O18}<1\,$barn$=10^{-24}$\,cm$^2$\\
Extinction length&&$\lambda\equiv 1/(\rho_{O18}\sigma_{O18})\,>\,84\,$cm\\
\hline
 \end{tabular}
 \end{table}

\section{Maximum energy loss of a proton scattering from an electron 
and the largest scattering angle}

The proton beam loses energy traversing tissue largely by Coulomb interactions with material electrons. Consider the Coulomb scattering of a relativistic proton with electrons in tissue.  As electrons bound to atoms and molecules have energies much lower than
the initial protons in a proton beam, they will recoil much like free electrons until the beam proton energies are much below $1\,$MeV. In this quasi-free region (ahead of the Bragg peak), the upper limit to the scattered electron's energy is determined by energy-momentum conservation.  Denote the
energy and momentum of the initial and final particles as
$(E_1,\vec{p}_1),(m,0),(E_3,\vec{p}_3),(\sqrt{k^2+m^2},\vec{k})$ and the mass of the proton as $M$. 

\subsection{Maximum delta-electron energy}

We will find the maximum electron energy after a collision with a beam proton.  Label the incoming particles with the indices $1, 2$, and outgoing particles $3, 4$.  We will distinguish the electron magnitude of its 3-mommentum with the symbol $k$. 

Conservation of energy and momentum reads
\begin{eqnarray}
E_{1}+E_{2} &=&E_{3}+E_{4} \\
\overrightarrow{p}_{1}+\overrightarrow{p}_{2} &=&\overrightarrow{p}_{3}+%
\overrightarrow{p}_{4}
\end{eqnarray}

The electron receives its greatest energy for head-on collisions.  In the `Lab' frame, $\overrightarrow{p}_{2}=\overrightarrow{0}$. We will use $\vec{k}=\vec{p}_4$ for the electron. Then
\begin{equation}
p_{1}=p_{3}+k
\end{equation}
and
\begin{equation}
\sqrt{p_{1}^{2}+M^{2}}+m=\sqrt{\left( p_{1}-k\right) ^{2}+M^{2}}+\sqrt{%
k^{2}+m^{2}}
\end{equation}
Solving for $k$, and dropping the $k=0$ solution, we have
\begin{equation}
k_{e}^{max } =2m\frac{\sqrt{p_{1}^{2}+M^{2}}+m}{M^{2}+m^{2}+2m\sqrt{%
p_{1}^{2}+M^{2}}}p_{1} \ .
\end{equation}

\begin{figure}[thpb]
\caption{Electron range in water. (\citet{plante2009cross})}
\center
\includegraphics[width=0.7\columnwidth]{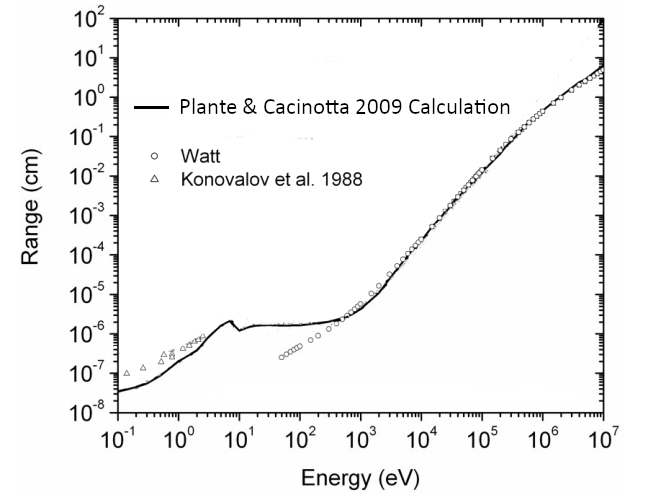}
\label{fig-e-range}
\end{figure}

There follows the maximum electron total energy
\begin{eqnarray}
E_{e}^{max } &=&\sqrt{\left( 2m\frac{\sqrt{p_{1}^{2}+M^{2}}+m}{%
M^{2}+m^{2}+2m\sqrt{p_{1}^{2}+M^{2}}}p_{1}\right) ^{2}+m^{2}} \\
&=&m\sqrt{1+8\frac{\left( 1+\frac{m}{M}+\frac{K_{1}}{M}\right) ^{2}}{\left(
\left( 1+\frac{m}{M}\right) ^{2}+2\frac{m}{M}\frac{K_{1}}{M}\right) ^{2}}%
\left( 1+\frac{1}{2}\frac{K_{1}}{M}\right) \frac{K_{1}}{M}} \ .
\label{eq-emaxT}
\end{eqnarray}
The maximum energy loss of a proton by electron scattering will by the maximum electron kinetic energy given to the electron. For $85\,$MeV incoming protons, Eq.\,(\ref{eq-emaxT})
gives
\begin{equation}
K_{e}^{max }=E_e^{max}-m=193.4\,\text{keV}
\end{equation}
The range of such electrons is about $0.2\,$ mm (see the graph in Fig.\,\ref{fig-e-range} which was presented by \citet{plante2009cross}). Most delta electrons
will have kinetic energy much less than $K_{e}^{max}$, and so a shorter
range, making their excursion from the proton beam track measurable in the tenths of a millimeter.

\subsection{Maximum beam proton deflection in scattering by electrons}
\label{app-max-theta}

Now let's find the maximum proton deflection angle, $\theta ^{\max }$ by one collision with an electron.
We have, from energy-momentum conservation
\begin{eqnarray}
E_{1}+m &=&E_{3}+E_{e} \\
\overrightarrow{p}_{1} &=&\overrightarrow{p}_{3}+\overrightarrow{k}
\end{eqnarray}
so 
\begin{eqnarray}
p_{1}^{2}+p_{3}^{2}-2p_{1}p_{3}\cos \theta &=&k^{2} \\
\cos \theta &=&\frac{p_{1}^{2}+p_{3}^{2}-k^{2}}{2p_{1}p_{3}}
\end{eqnarray}
\begin{eqnarray}
E_{3}^{2} &=&\left( E_{1}+m-E_{e}\right) ^{2} \\
p_{3} &=&\sqrt{E_{3}^{2}-M^{2}} \\
&=&\sqrt{\left( E_{1}+m-E_{e}\right) ^{2}-M^{2}}
\end{eqnarray}
Thus 
\begin{equation}
\cos \theta =\frac{E_{1}^{2}-M^{2}+\left( E_{1}+m-E_{e}\right)
^{2}-M^{2}-\left( E_{e}^{2}-m^{2}\right) }{2\sqrt{E_{1}^{2}-M^{2}}\sqrt{%
\left( E_{1}+m-E_{e}\right) ^{2}-M^{2}}} \ .
\label{eq-cos}
\end{equation}

This cosine function is  $1$ ($\theta=0$) when there is no scattering ($E_e=m$) or at
\begin{equation}
K_e^{\theta=0}=\frac{mp_1^2}{M^2+m(E_1+(1/2)m)}
\end{equation}
with scattering.
It drops below one as $K_e$ increases. 

As an aside, if $m=M$ (as for proton-proton scattering), or $m>M$
the cosine function drops to zero at
\begin{equation}
K_e^{\theta=90^o}=\frac{p_1^2}{(E_1+m)} \ .
\end{equation}

If $m<M$, $\cos\theta$ rises back to one at a
finite electron kinetic energy given by
\begin{eqnarray}
K_e^{\theta=90^o}&=&2\frac{2M+K_1}{M^2+2mM+m^2+2mK_1}mK_1 \ , \\
&=&4m\frac{1+\frac{1}{2}\frac{K_1}{M}}{\left(1+\frac{m}{M}\right)^2+2\frac{m}{M}\frac{K_1}{M}}\frac{K_1}{M} \ .
\end{eqnarray}

We can find the energy where the greatest deflection
occurs by finding where the slope of $\cos \theta $ vanishes with changing $%
E_{e}.$ Setting 
\begin{equation}
\frac{d}{dE_{e}}\left( \frac{E_{1}^{2}-M^{2}+\left( E_{1}+m-E_{e}\right)
^{2}-M^{2}-\left( E_{e}^{2}-m^{2}\right) }{2\sqrt{E_{1}^{2}-M^{2}}\sqrt{%
\left( E_{1}+m-E_{e}\right) ^{2}-M^{2}}}\right) =0 \ ,
\end{equation}
or
\begin{equation}
E_{1}m^{2}+E_{1}^{2}m-M^{2}E_{e}-E_{1}mE_{e} = 0 \ ,
\end{equation}
which gives 
\begin{equation}
E_{e}^{\theta -\max } = m\frac{m+E_{1}}{E_{1}m+M^{2}}E_{1} \ ,
\label{eq-E-theta}
\end{equation}
so the kinetic energy of the electrons when $\theta$ is maximum as
\begin{eqnarray}
K_{e}^{\theta -\max } &=&m\frac{m+M+K_{1}}{\left( M+K_{1}\right) m+M^{2}}%
\left( M+K_{1}\right) -m \\
 &=&m\frac{2M+K_{1}}{M^{2}+Mm+mK_{1}}K_{1} \ . \\
&=& 2m\frac{1+\frac{1}{2}\frac{K_1}{M}}{1+\frac{m}{M}(1+\frac{K_1}{M}K_1)}
\end{eqnarray}

For an $85$\thinspace MeV beam, this is
\begin{eqnarray}
K_{e}^{\theta -\max } &=&2\times \left( \frac{0.511}{938}\right) \times 
\frac{1+0.5\times \frac{85}{938}}{(1+\left( \frac{0.511}{938.}\right) \times
\left( 1+\frac{85}{938}\right) }\times 85 \\
&=&0.0968\text{\thinspace MeV}=96.8\,\,\text{keV}
\end{eqnarray}

Inserting $E_e^{\theta-max}$ from Eq.\,(\ref{eq-E-theta}) in the formula Eq.\,(\ref{eq-cos}) for $ \cos \theta $ gives {\bf a remarkably simple result}, and one that does not depend on the proton beam
energy: 
\begin{eqnarray}
\cos \theta ^{\max } &=&\sqrt{1-\left( \frac{m}{M}\right) ^{2}} \ , \\
\sin \theta ^{\max } &=&\left( \frac{m}{M}\right) \ , \\
\theta ^{\max } &=&\arcsin \left( \frac{m}{M}\right) =\frac{180}{\pi }\times
\arcsin \left( \frac{0.511}{938}\right) =3.12\times 10^{-2}\,\deg \ .
\end{eqnarray}
Each time a beam proton interacts with a medium electron, the proton cannot be
deflected more than $0.0312$ degrees.

\begin{figure}[thpb]
\caption{``A curious fact" (\citet{gottschalk2018})}
\center
\includegraphics[width=0.6\columnwidth]{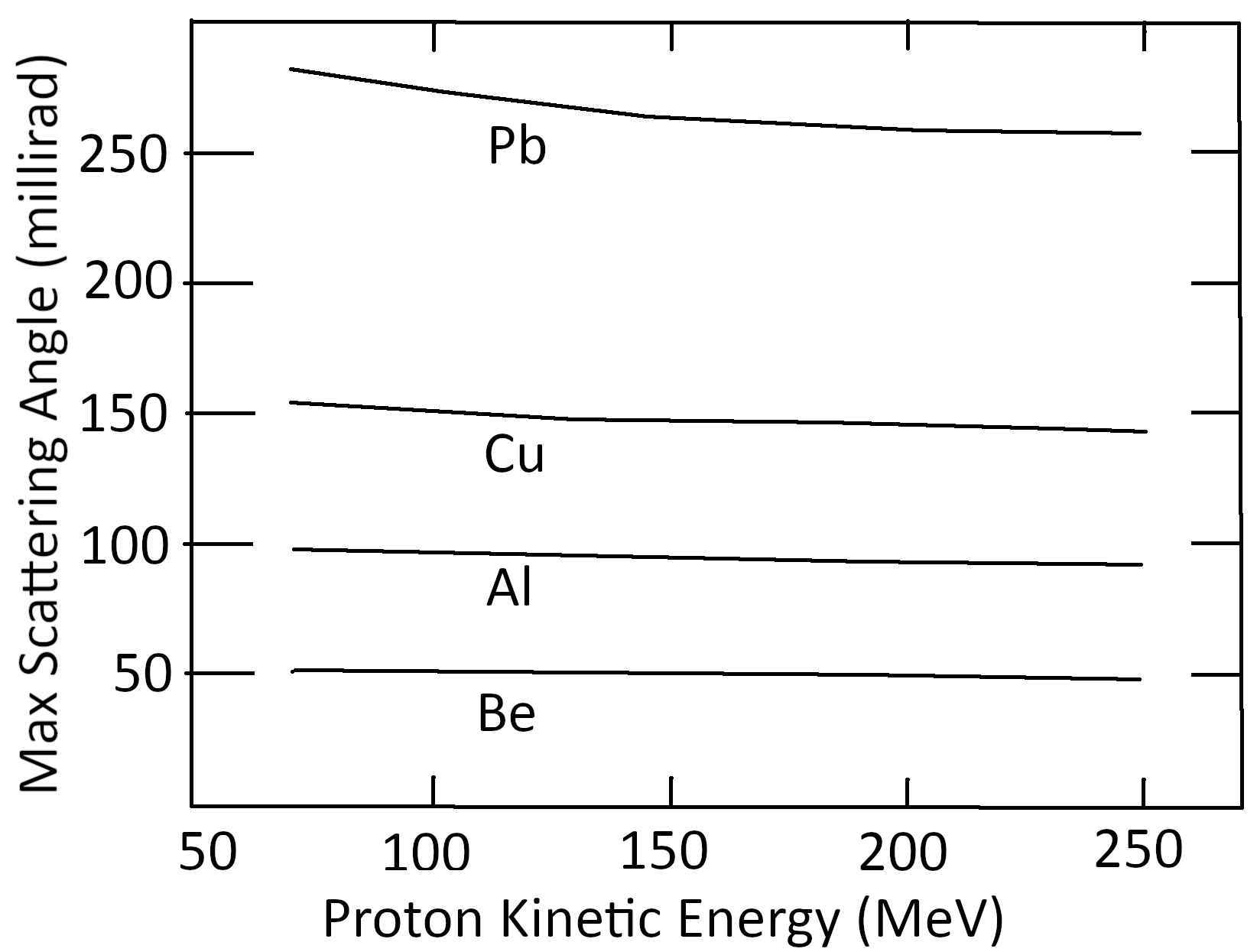}
\label{fig-curious}
\vspace{0.5in}
\end{figure}

The experimental independence of the maximum scattering angle with beam proton energy
was noted by Andy Koehler as reported by Bernard Gottschalk as a `curious fact' in his lecture series \cite{gottschalk2018lec} and depicted as Fig.\,\ref{fig-curious}.
We now see that the energy independence of the maximum scattering angle (even after
several scatterings) follows from kinematics.

\section{Energy and momentum of emitted particles after a particle beta-decay}

To simplify indices, we will use here (0,1,2,3) for the parent, daughter, beta, and neutrino.  A notation on top of an energy or momentum symbol indicates a frame of reference.  Zero will be used for the frame in which the parent nucleus is at rest. (This is `the center of momentum (cm) frame'.)

\vspace{0.1in}

Energy-momentum conservation gives (for the 4-vectors)
\begin{equation}
p_{0}=p_{1}+p_{2}+p_{3} \ .
\end{equation}
In the rest frame of the parent particle, the three vector-momenta must lie
in a plane. The orientation of this plane determines two of the independent
kinematic variables measurable in the reaction. Orientation of one of the
outgoing momenta along some axis in the cm-reaction plane determines
another. Momentum conservation means one of the three momenta can be written
in terms of the other two.  Energy conservation gives the magnitude of one
of the outgoing momenta in terms of  the other two. Of the nine variables
in the three momenta, the constraints mention leaves 9-2-1-3-1=2 variables
not determined by kinematics. These two might be taken as the angle between
particles (1,2) and the energy of particle 2. A more universal choice, a
choice that does not depend on a selection of reference frame, is to use
kinematical variables that are relativistic invariants. These are often
taken as a pair of Mandelstam variables, defined by
\begin{eqnarray}
s_{1} &=&\left( p_{2}+p_{3}\right) ^{2}\,, \\
s_{2} &=&\left( p_{3}+p_{1}\right) ^{2}\,, \\
s_{3} &=&\left( p_{1}+p_{2}\right) ^{2}\,.
\end{eqnarray}
These three are not independent of each other, as can be seen from
\begin{equation}
s_{1}+s_{2}+s_{3}=m_0^2+m_1^2+m_2^2+m_3^2 \ .
\label{sum-s}
\end{equation}

To preserve symmetry among the three variables in the plot of the
kinematically allowed values of $(s_{1},s_{2},s_{3}),$ Dalitz first plotted
the possible values of $(s_1,s_2,s_3)$ in three dimensions.  The constraint Eq. (\ref{sum-s})
forces the values to lie in a plane oriented with a normal having positive
values on each axis.  This plane cuts the (1,2), (2,3), and (3,1)
axis-planes, producing a triangular surface. \ The Dalitz plot is made on
that surface.  The density of values as points within the triangle are made
proportional to the physical probabilities for detecting particles produced
with the kinematics of the point.

Now the question arises: What are the kinematical limits within the Dalitz
triangle, or, equivalently, what are the limits on the measured values of
momentum and energy of the outgoing particles. Such physical limits come
from the constraints that the kinetic energies of the particles cannot be
negative ($E_{i}\geq m_{i}$), and that the angles between the momenta of the
outgoing particles must be physical ($-1\leq \cos \theta \leq 1$).

To get an upper limit on $s_{1}$, we note that $\left( p_{0}\cdot
p_{1}\right) ,$ in the frame of reference of the parent particle, becomes
\begin{equation}
p_{0}\cdot p_{1}=m_{0}\overset{0}{E}_{1}
\end{equation}
But
\begin{equation}
s_{1}=\left( p_{2}+p_{3}\right) ^{2}=\left( p_{0}-p_{1}\right)
^{2}=m_{0}^{2}+m_{1}^{2}-2p_{0}\cdot p_{1}
\end{equation}
giving
\begin{equation}
\overset{0}{E}_{1}=\frac{1}{2m_{0}}\left( m_{0}^{2}+m_{1}^{2}-s_{1}\right) 
\end{equation}
From $\overset{0}{E}_{1}\geq m_{1},$ we have
\begin{equation}
s_{1}\leq \left( m_{0}-m_{1}\right) ^{2}
\end{equation}
To get a lower limit on $s_{1},$ pick the rest frame for the pair of
particles (2,3) (the `Jackson' frame),  so that    
\begin{equation}
\overset{J}{\overrightarrow{p}_{2}}=-\overset{J}{\overrightarrow{p}%
_{3}}
\end{equation}
Then 
\begin{equation}
s_{1}=\left( p_{2}+p_{3}\right) ^{2}=\left( \overset{J}{E_{2}}+\overset{%
J}{E_{3}}\right) ^{2}\geq \left( m_{2}+m_{3}\right) ^{2}
\end{equation}
So, for physically realizable energies, (by cyclically permuting of indices)
\begin{equation}
\left( m_{2}+m_{3}\right) ^{2}\leq s_{1}\leq \left( m_{0}-m_{1}\right) ^{2}
\end{equation}
\begin{equation}
\left( m_{3}+m_{1}\right) ^{2}\leq s_{2}\leq \left( m_{0}-m_{2}\right) ^{2}
\end{equation}
\begin{equation}
\left( m_{1}+m_{2}\right) ^{2}\leq s_{3}\leq \left( m_{0}-m_{3}\right) ^{2}
\end{equation}
When angular constraints are imposed, the $s^{\prime }$s may not be allowed
to reach the above limits. In the `Jackson' frame for particles 2 and
3', in which the momentum of particles 2 and 3\ satisfy $\overset{J}{%
\overrightarrow{p}_{2}}=\overset{J}{-\overrightarrow{p}_{3}}$. Furthermore:
\begin{eqnarray}
s_{1} &=&\left( p_{0}-p_{1}\right) ^{2} \\
&=&\left( E_{0}-E_{1}\right) ^{2}-\left( \overrightarrow{p}_{0}-%
\overrightarrow{p}_{1}\right) ^{2} \\
s_{1} &=&\left( E_{0}-E_{1}\right) ^{2}-\left( \overrightarrow{p}_{2}+%
\overrightarrow{p}_{3}\right) ^{2} \\
&=&\left( \overset{J}{E_{0}}-\overset{J}{E_{1}}\right) ^{2} \\
s_{1} &=&\left( \sqrt{\left( \overset{J}{\overrightarrow{p}_{1}}\right)
^{2}+m_{0}^{2}}-\sqrt{\left( \overset{J}{\overrightarrow{p}_{1}}\right)
^{2}+m_{1}^{2}}\right) ^{2} \ .
\end{eqnarray}
As a result,
\begin{equation}
\left( \overset{J}{\overrightarrow{p}_{1}}\right) ^{2} =\frac{1}{4}\frac{%
\left( \left( m_{0}^{2}+m_{1}^{2}\right) ^{2}-s_{1}^{2}\right) \left( \left(
m_{0}^{2}-m_{1}^{2}\right) ^{2}-s_{1}^{2}\right) }{s_{1}^{2}} \ .
\end{equation}
Now from
\begin{eqnarray}
s_{1} &=&\left( p_{2}+p_{3}\right) ^{2}=\left( \overset{J}{E_{2}}+\overset{J%
}{E_{3}}\right) ^{2} \\
&=&\left( \sqrt{\left( \overset{J}{\overrightarrow{p}_{2}}\right)
^{2}+m_{2}^{2}}+\sqrt{\left( \overset{J}{\overrightarrow{p}_{2}}\right)
^{2}+m_{3}^{2}}\right) ^{2} \ ,
\end{eqnarray}
we can solve for the momentum squared of particle 2 in the Jackson frame gives
\begin{equation}
\left( \overset{J}{\overrightarrow{p}_{2}}\right) ^{2}=\frac{%
m_{2}^{4}+m_{3}^{4}+s_{1}^{2}-2s_{1}m_{2}^{2}-2s_{1}m_{3}^{2}-2m_{2}^{2}m_{3}^{2}%
}{4s_{1}} \ ,
\end{equation}
which is also the momentum squared for particle 3 in the Jackson frame. \ We
now consider
\begin{eqnarray}
s_{2} &=&\left( p_{3}+p_{1}\right) ^{2} \\
&=&m_{3}^{2}+m_{1}^{2}+2\left( p_{3}\cdot p_{1}\right)  \\
&=&m_{3}^{2}+m_{1}^{2}+2\left( \overset{J}{E_{3}}\overset{J}{E_{1}}-\left| 
\overset{J}{\overrightarrow{p}_{3}}\right| \left| \overset{J}{%
\overrightarrow{p}_{1}}\right| \cos \theta \right) \ .
\end{eqnarray}
Since, as we have shown, $\left| \overset{J}{\overrightarrow{p}_{3}}\right| $
and $\left| \overset{J}{\overrightarrow{p}_{1}}\right| $ can be expressed in
terms of $s_{1}$ (without $s_{2}$), for fixed $s_{1}$, we have
\begin{equation}
m_{3}^{2}+m_{1}^{2}+2\left( \overset{J}{E_{3}}\overset{J}{E_{1}}-\left| 
\overset{J}{\overrightarrow{p}_{3}}\right| \left| \overset{J}{%
\overrightarrow{p}_{1}}\right| \right) \leq s_{2}\leq
m_{3}^{2}+m_{1}^{2}+2\left( \overset{J}{E_{3}}\overset{J}{E_{1}}+\left| 
\overset{J}{\overrightarrow{p}_{3}}\right| \left| \overset{J}{%
\overrightarrow{p}_{1}}\right| \right) 
\end{equation}
with the lower limit in the case in which $\overset{J}{\overrightarrow{p}_{1}%
}$ is in the same direction as $\overset{J}{\overrightarrow{p}_{3}}$ and the
upper limit in the case in which  $\overset{J}{\overrightarrow{p}_{1}}$ is
in the opposite direction as $\overset{J}{\overrightarrow{p}_{3}}$. These
limits determine the boundary in the Dalitz plot in the $\left(
s_{1},s_{2}\right) $ plane.

To get the maximum energy in the frame in which the parent particle is at rest
(center-of-momentum frame), we note that
\begin{equation}
s_{1}=\left( p_{0}-p_{1}\right) ^{2}=m_{0}^{2}+m_{1}^{2}-2m_{0}\overset{0}{E}%
_{1}
\end{equation}
implies that $\overset{0}{E}_1$ is determined by just one of the two independent Mandelstam variables, and that $\overset{0}{E}_{1}$ reaches a maximum when $s_{1}$ is a minimum.
Solving for $\overset{0}{E}_{1},$ we have
\begin{equation}
\overset{0}{E}_{1}=\frac{m_{0}^{2}+m_{1}^{2}-s_{1}}{2m_{0}} \ .
\end{equation}
Using the min of $s_{1}$ above,
\begin{equation}
\overset{0}{E}_{1}^{\max }=\frac{m_{0}^{2}+m_{1}^{2}-\left(
m_{2}+m_{3}\right) ^{2}}{2m_{0}} \ .
\label{eq-emaxS}
\end{equation}

It follows that
\begin{eqnarray}
p_{1}^{0\max } &=&\sqrt{\left( \frac{m_{0}^{2}+m_{1}^{2}-\left(
m_{2}+m_{3}\right) ^{2}}{2m_{0}}\right) ^{2}-m_{1}^{2}} \\
&=&\frac{1}{2m_{0}}\sqrt{\left( m_{0}^{2}-\left( m_{1}-m_{2}-m_{3}\right)
^{2}\right) \left( m_{0}^{2}-\left( m_{1}+m_{2}+m_{3}\right) ^{2}\right) } \ .
\end{eqnarray}

By cyclic permutation,
\begin{equation}
\overset{0}{E}_{2}^{\max }=\frac{m_{0}^{2}+m_{2}^{2}-\left(
m_{3}+m_{1}\right) ^{2}}{2m_{0}} \ ,
\end{equation}
and
\begin{equation}
\overset{0}{E}_{3}^{\max }=\frac{m_{0}^{2}+m_{3}^{2}-\left(
m_{1}+m_{2}\right) ^{2}}{2m_{0}} \ .
\end{equation}

\section{Reaction `Q' values and threshold energy in laboratory frame}
\label{app-Q}

Consider the reaction 
\begin{equation}
[1]+[2]\rightarrow [3]+[4]+[5]+\cdots \ ,
\end{equation}
where $[i]$ is particle labeled by index $i$. 

The ``Q'' value for this reaction is defined to be the energy released because of mass differences before and after the reaction:
\begin{equation}
Q=m_1+m_2-\sum_f m_f \ ,
\end{equation}
the sum taken over the produced (final) particles.

To find the threshold energy of particle $[1]$ in the ``lab'' frame (where $\vec{p}_2=0$),
consider the Mandelstam variable $s=(p_1+p_2)^2=E_{cm}^2$, 
the total energy squared in the center-of-momentum frame of reference.
From energy-momentum conservation, $s=(\sum_f p_f)^2$ (sum is over final particles), so we see that $\sqrt{s}\ge \sum_f m_f$.
Expressed in the laboratory frame variables, $s=m_1^2+m_2^2+2m_2E_1^{lab}$. Thus
\begin{equation}
E_1^{lab}\ge \frac{(\sum_f m_f)^2-m_1^2-m_2^2}{2m_2} \ ,
\end{equation}
making the threshold kinetic energy of the incoming particle satisfy
\begin{equation}
K_1^{th}\equiv E_1^{th}-m_1\ge \frac{(\sum_f m_f)^2-(m_1+m_2)^2}{2m_2} \ .
\end{equation}
In terms of $Q$,
\begin{equation}
K_1^{th}\ge - \frac{\sum_im_i}{2m_2}\, Q \ ,
\end{equation}
where now the sum is over all particles, in and out.
If the reaction is exoergic, $Q>0$ and then $K_1^{th}\ge 0$. If endoergic, $Q<0$,
the threshold kinetic energy for the incoming particle will be greater than zero, as
given above.

\section{Maximum energy of the recoiling O18 after beta decay of F18}
\label{app-recoil}

The maximum oxygen recoil kinetic energy, from Eq.\,(\ref{eq-emaxS}),
\begin{eqnarray}
E_{O} &=&\frac{m_{F-bare}^{2}+m_{O-bare}^{2}-m_{e}^{2}}{2m_{F-bare}} \\
K_{O} &=&\frac{m_{F-bare}^{2}+m_{O-bare}^{2}-m_{e}^{2}}{2m_{F-bare}}%
-m_{O-bare} \\
 &=& \frac{(m_{F-bare}-m_{O-bare})^2-m_e^2}{2m_{F-bare}}\\
&=&3.\, 132\,7\times 10^{-5}\,\text{MeV}=31.3\,\text{eV} \ .
\end{eqnarray}
The energy needed to cause a double-strand break in DNA is about $25\,$eV, so the recoil kinetic energy of the O18 is sufficient to break both strands of its DNA, but this breakage depends on the O18 recoil direction. The electron, on the other hand, carries, on average, about $250$ keV, which can leave a trail of ions by scattering from molecular electrons along its path.  For F18, the path is relatively short (average $0.24\,$mm in soft tissue (see Table \ref{beta-emit})).

\section{Fate of the positron after the beta decay of F18}
\label{sec-fate}

As shown above, in the particle-decay reaction $\mathcal{P}_0\rightarrow \mathcal{P}_1+\mathcal{P}_2+\mathcal{P}_3$, the maximum recoil energy of the particle $\mathcal{P}_2$ after the parent decays can be found from energy and momentum conservation, giving
\begin{equation}
E_{2}=\frac{m_{0}^{2}+m_{2}^{2}-\left(m_{3}+m_{1}\right) ^{2}}{2m_{0}} \ .
\end{equation}
For F18 decay, the positron maximum total energy and kinetic energy are therefore
\begin{eqnarray}
E_{e} &=&\frac{m_{F-bare}^{2}+m_{e}^{2}-m_{O-bare}^{2}}{2m_{F-bare}} \\
&=&1.1451 \ \text{MeV}\ ,\\
K_{e} &=&E_{e}-m_{e}=1.1451-0.511=0.6341 \ \text{MeV}\ .
\end{eqnarray}
$K_e$ agrees with the experimental number $0.634\,$MeV, given in Table \ref{beta-emit}.

The positron energy is typically more than 200 times
greater than the highest ionization energy of atoms and molecules in tissue (see Table \ref{tab-m-dense}). As the fast-moving positron (starting at $89\%$ of the speed of light) travels through tissue, it inelastically scatters from bound electrons and nuclei and slows down.  Since the positron-electron cross section depends inversely on
the center-of-mass energy squared, the positron loses most of its energy in tissue
at the end of its track, just as ions beams do. We should 
expect the emitted positron to scatter among the nearby molecular charges,
leading to molecular excitations, ionizations, and photons, 
before slowing to thermal energies.  It can then be directly annihilated or captured by an electron. In water at $20^o$C, the positron has about a $64$\% chance of undergoing direct
annihilation (\cite{colombino1971}). 
If captured, positronium forms, either in a spin single state (ortho) or triplet state (para).  With no preferred direction of the positron spin, $25$\% will be ortho-positronium, although interaction with nearby molecules can flip one of the spins. If
the positronium is created in an excited state with angular momentum, it will release UV and visible-light
photons with discrete energies half those of atomic hydrogen, until the $e^+e^-$ pair reach an $l=0$ state (no angular momentum).
At that point they annihilate each other. If the positronium is in a singlet spin state, the annihilation has a lifetime of $0.12$ ns and almost always two $0.511\,$MeV photons are emitted in opposite directions. The spin triple state annihilates with a lifetime of $142$ ns, with almost always the
emission of three gamma rays.

\section{Threshold behavior of nuclear reaction cross sections}
\label{sec-gamow}

In 1948, Eugene Wigner showed (\cite{wigner1948}) that under quite general circumstances, the low energy threshold behavior of nuclear-reaction cross-sections have a definite dependence on the relative momentum $\hbar k$ between the two incoming particles, given by
\begin{eqnarray}
\sigma \propto \frac{1}{k^2}\exp{\left(-2\pi \mu c Z_1Z_2 \alpha/(\hbar k)\right)} &\ \ \ \ \text{for (+)(+) or (-)(-) charges} \label{pp}\\
\sigma \propto \frac{1}{k^2} & \text{ for (+)(-) charges} \ ,
\end{eqnarray} 
where $\mu$ is the reduced mass of the two particles, having charge numbers $Z_1$ and $Z_2$, respectively, $c$ is the speed of light, and $\alpha$ is the fine-structure constant.
A nuclear particle with a positive charge sent toward another one will be inhibited from merging because of the Coulomb repulsive force acting. The effect is expressed by the exponential factor in Eq.\,(\ref{pp}), used by George Gamow in 1928 to explain alpha-particle decay of nuclei, and is now called the Gamow factor.

Gamow realized \cite{gamow1928decay} that if two particles are in a bound state held together by a strong short-range nuclear force and repelled by a Coulomb force, they may still escape from each other by `tunneling' through the potential barrier created by the two forces.  With this quantum idea, he was able to explain the wide range of lifetimes of alpha-particle decay of radioactive nuclei. 

Inversely, the probability of two nuclear particles, with masses $m_1$, $m_2$ and charges $Z_1\left|e\right|$, $Z_2\left|e\right|$, approaching each other and then overcome the Coulomb repulsion, is:
\begin{equation}
P_g(K)=\exp{\left(-\sqrt{\frac{K_g}{K}}\right)}
\label{eq-prob}
\end{equation}
where $K$ is their total  kinetic energy, $K_g=2\mu c^2(\pi \alpha Z_1Z_2)^2$, with
$c$ the speed of light, $\mu\equiv m_1m_2/(m_1+m_2)$ (i.e. the reduced mass) and $\alpha\equiv e^2/(\hbar c)$ (the fine-structure constant). Note that as the kinetic energy $K$ goes to zero, the probability of penetration goes to zero. Eq.\,(\ref{eq-prob}) is a non-relativistic expression. Yoon and Wong \cite{yoon2000} have given the relativistic form for the Gamow factor.

By factoring out the strongly energy dependent factors $P_g$ and $1/K$ from the cross section,
the remaining `astrophysical factor $S(K)$' will have a more gentle low-energy behavior:
\begin{equation}
\sigma = \frac{1}{K}P_g(K)S(K) \ .
\end{equation}
The $S(K)$ is proportional to the nuclear transition probability, so it contains all the nuclear physics of the process. Apart from nuclear resonances, one finds that $S(K)$ is quite flat for $K$ in the eV to keV range.

\section{Proton energy at given penetration depth}
\label{sec-K-z}

As protons from a PT beam enters tissue, a variety of mechanisms of interaction cause the protons to
lose kinetic energy $K$, and to reduce their number (dropping beam fluence). The ionization of molecular electrons dominates, particularly above a few MeV. The inelastic collision with nuclei, although important because of the creation of nuclear fragments and because such collisions spread the beam, contributes only a small fraction to the proton energy-loss because of the small cross sections of nuclei.  
The energy loss of the ions in transversing through uniform tissue of a given short length (Linear Energy Transfer, LET)  follows closely to the Bethe-Bloch relation,
with corrections, shown in Eq.\,(\ref{eq-BB}): (See \citet{bethe1930},\citet{fano1947}, \citet{ziegler1999}.)
\begin{equation}
\frac{dK}{dz}=-4\pi n_e \alpha^2 z^2\frac{(\hbar c)^2}{m_ec^2}\frac{1}{\beta^2}\left(\ln{\left(\frac{2m_ec^2}{I}\right)}+\ln{\left(\frac{\beta^2}{1-\beta^2}\right)}-\beta^2-\delta-\frac{C}{z}\right) \ .
\label{eq-BB}
\end{equation}
Here, $n_e$ is the material electron density, $z$ the number of charges on the beam ion,
$m_e$ the electron mass, $I$ the average ionization energy in the material, and $\beta$ is
the velocity of the beam ions divided by the speed of light. The term $\delta$ is a material electron density correction due
to polarization of surround material as a relativistic ion passes, whose electric field spreads perpendicular to the ions velocity and shrinks parallel to that velocity for larger speeds. The term $C/z$ is a ``shell'' correction needed when the beam particle
slows to speed comparable to the speed of the bound electrons in the tissue molecules.  For protons in energy range $1\,$MeV to $100\,$MeV, the shell correction can be as much as 6\% \cite{ziegler1999}. In the derivation of Eq.\,(\ref{eq-BB}), the discrete nature of
protons interacting with a medium is smoothed (``Continuous Slow-Down Approximation'', or CSDA for short), since, for most of its interaction with electrons, small effects rapidly
occurring take place.  These days, the discrete events can be handled by Monte Carlo methods (e.g., using the GEANT4 code \cite{collaboration2003geant4}, \cite{allison2006geant4}, \cite{allison2016recent}). However, analytic tools involving continuous changes often can
quite accurately account for discrete processes and many times the analytic arguments are simpler to understand and handle.

To integrate the LET relation, we can use the relativistic connection between the kinetic energy and velocity of the protons. We have, in units with $c=1$ and with $M$ the mass of the proton,
\begin{equation}
K=\frac{M}{\sqrt{1-\beta^2}}-M
\end{equation}
and, inversely
\begin{equation}
\beta=\sqrt{1-(1+K/M)^{-2}}=\frac{\sqrt{K(K+2M)}}{K+M} \ ,
\end{equation}
from which we can connect $dK$ to $d(\beta^2)$, facilitating the integration needed in Eq.\,(\ref{eq-BB}) to find 
$z=z(K)$, which is an implicit solution for $K=K(z)$. The connection is
\begin{equation}
dK=\frac{M}{2}(1-\beta^2)^{-3/2}d(\beta^2) \ .
\end{equation}

When the proton speed becomes very small ($\beta \ll 1$), we can use 
$\underset{v\rightarrow 0}{lim}\,(\ln{\beta^2})/\beta^2=0$, so that we can see that
the large value of the LET (causing  the ``Bragg peak'') comes from the first term with the inverse of $\beta^2$ factor. This term generates an infinite peak for zero proton speed. However, we should not expect the Bethe formula to work when
the protons have slowed to below the average ionization energy $I$, which is typically found under $100\,$eV (see Table \ref{tab-m-dense}). At proton kinetic energy of $K=100\,$eV, $\beta \approx 0.014$. In turn,
this value for $\beta$ is far above the thermal energy at body temperature $T=310\,$Kelvin, which is $\beta\approx \sqrt{3k_BT/(2mc^2})=1.86\times 10^{-4}$. If we take $I=100\,$eV, then the proton speed that makes
Bethe's $dK/dz$ become negative is $\beta\approx 0.0067,$ which is below
the range of validity of the formula.

To simplify integrating, one can use $\beta^2 < 1$, so an
expansion of the second term in powers of $\beta^2$ is justified:
\begin{equation}
\frac{dK}{dz}=-4\pi n_e \alpha^2 z^2\frac{(\hbar c)^2}{m_ec^2}\frac{1}{\beta^2}\left(\ln{\left(\frac{2m_ec^2}{I}\right)}-\ln{(1/\beta^2)}
+\frac{1}{2}\beta^4+\frac{1}{3}\beta^6+\frac{1}{4}\beta^8+\cdots\right) \ .
\end{equation}
For an initial proton energy of $K=100\,$MeV, $\beta=0.428$, $\beta^4=0.0336$, while $\ln{(\beta^2)}=-0.417$. The terms beyond the $\ln{(\beta^2)}$ have an even smaller contribution, and can be dropped.

\begin{figure}[!ht]
\caption{Bragg peak (\citet[p.9]{grun2014impact})}
\center
\includegraphics[width=0.6\columnwidth]{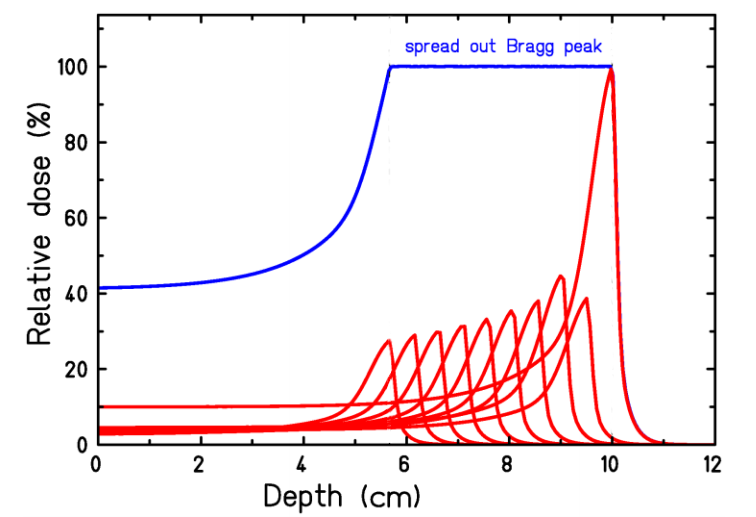}
\label{fig-5}
\end{figure}

The proton beam Bragg peak is shown in Figure \ref{fig-5}. By including a set of different initial proton energies in the incoming beam, the different ranges of the Bragg peaks can be made to produce a spread out Bragg peak (SOBP), designed to cover the width of a tumor. To reduce exposure of healthy tissue in front of the tumor for a given dose to the tumor and to work around structures needing avoidance, the beam angle can be rotated to a set of angles.

The dose delivered along the proton beam path is well represented by Bortfield's analytic expression derived from the Bethe-Block relation: (\citet[eq.26]{bortfeld1997}, \citet[eq.38]{newhauser2015})
\begin{equation}
D(z)=\Phi_0\frac{e^{-(z-R_0)^2/(4\sigma_B^2)}\Gamma(q_B+1)}{\sqrt{2\pi}\rho (1+\beta_B R)}\left(\frac{\sigma_B}{\alpha_B}\right)^q\left[\frac{1}{\sigma_B}D_{-q_B}\left((z-R)/\sigma_B\right)
+\left((q_B+\gamma_B) \beta_B + \frac{\varepsilon_B}{R}\right)D_{-q_B-1}\left((z-R)/\sigma_B\right)\right]\ ,
\label{eq-dose}
\end{equation}
where $z$ is the depth, $\Phi_0$ is the primary fluence, $R$ is the range of the proton beam, $\sigma_B$ is the standard deviation of the Gaussian spread of the proton depth, $\varepsilon$ is the fraction of low-energy proton fluence to the total proton fluence, and $D_y(x)$ is the parabolic cylinder function. The values of the material-dependent constants, $q_B$ and $\alpha_B$, are found by fitting them using the classical LET  Bragg-Kleeman (1905) formula: 
\begin{equation}
\frac{dK}{dz}=-q_B\frac{K^{1-1/q_B}}{\alpha_B} \ .
\label{eq-BK}
\end{equation}
to experiment. Of course, the Bragg-Kleeman formula is far simpler than the Bethe-Bloch result,
but it still works surprisingly  well.
(The fitted constant $q_B$ is $0.565$.  If $q_B=1/2$, then the parabolic cylinder functions in Eq.\,(\ref{eq-dose}) become a Gaussian times a Hermite polynomial.)
The parameters fitted by Bortfeld for a water target are given in Table \ref{tab-bort}. The standard deviation of the range spectrum for an almost mono-energetic beam is denoted $\sigma_{mono}$ and the standard deviation of the almost Gaussian part of the energy-spectrum at its peak $E_0$ is given by $\sigma_{E,0}$. These make up the full standard deviation of the range spectrum through
\begin{equation}
\sigma_B^2=\sigma^2_{mono}+\sigma^2_{K_0}\left(\frac{dR}{dK_0}\right)^2=\sigma^2_{mono}+\sigma^2_{E,0}\left(\frac{\alpha_B}{q_B}K_0^{1/q_B-1}\right)^2\ .
\end{equation}
\begin{table}
\caption{Dose-depth parameters for water (\citet{bortfeld1997})}
\vspace{0.1in}
\begin{tabular}{p{0.6 in}lc}
\hline
\hline
 &  Value & Units \\
\hline
$q_B$  & 0.565 & 1 \\
$\alpha_B$ & $0.0022$ & cm/MeV$^{1/q}$ \\
$\beta_B$ & $0.012$ & cm$^{-1}$ \\
$\gamma_B$ & 0.6 & 1 \\
$R$ & $\alpha_B K^{1/q}_0$ & cm \\
$\sigma_{mono}$ & $\beta'\,R^{3/2-q}_0$ & cm \\
$\beta^{\prime}$ & $0.012$ & cm$^{q-1/2}$ \\
$\sigma_{E,0}$ & $\approx 0.01\,K_0$ & MeV \\
$\varepsilon_B$ & $\approx 0.0-0.2$ & 1 \\
\hline
\hline
\end{tabular}
\label{tab-bort}
\end{table}

\begin{figure}[!ht]
\caption{Proton kinetic energy vs z}
\center
\includegraphics[width=0.7\columnwidth]{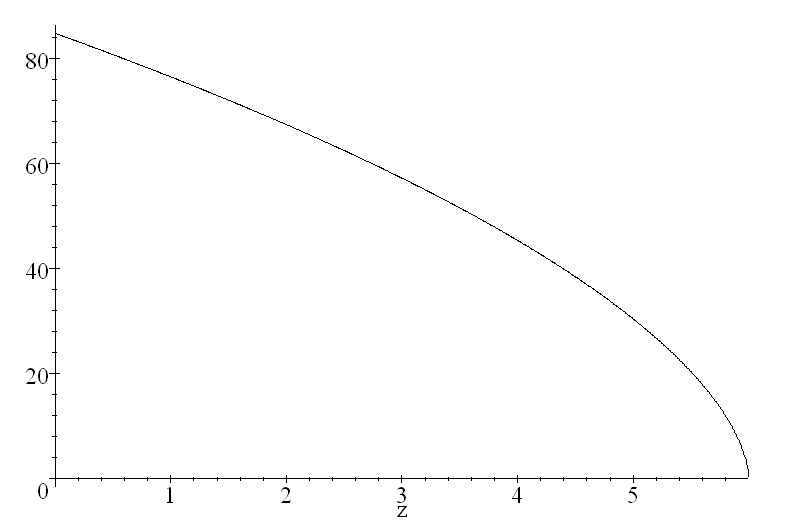}
\label{K-vs-z}
\end{figure}

An even better fit to the classical LET (Eq.\,(\ref{eq-BK})) is a generalization to relativistic energies (see \citet{ulmer2010}), taking a classical slow-down due to a frictional drag (`damping') proportional to the protons momentum $p_z$ along the beam axis to an inverse power:
\begin{equation}
\frac{dp_z}{d\tau}=-\eta /p_z^q ,
\label{eq-ud}
\end{equation}
where $\tau$ is the relativistic proper-time $d\tau=\sqrt{1-\beta^2}dt$.
Integrating Eq.\,(\ref{eq-ud}) once gives
\begin{equation}
p_z=p_o(1-\tau/\tau_R)^{1/(q+1)} \ ,
\label{eq-pr}
\end{equation}
where $p_o$ is the initial proton beam momentum and $\tau_R=p_0^{q+1}/((q+1)\eta)$ is the proper  time
to stop the protons in the tissue.  We can integrate Eq.\,(\ref{eq-pr}), to get the proton distance
in the tissue in terms of the proper time to reach that distance, i.e.
\begin{equation}
z=R\left[1-\left(1-\frac{\tau}{\tau_R}\right)^{\frac{q+2}{q+1}}\right] \ ,
\label{eq-z}
\end{equation}
where $R=((q+1)/(q+2))\tau_R p_0/M$ is the range of the proton. Now with Eq.\,(\ref{eq-pr}) and \,(\ref{eq-z}) we can express the momentum of the beam proton in terms of its distance of travel through the tissue:
\begin{equation}
p_z=p_o(1-z/R)^{1/(q+2)} \ .
\label{eq-pz}
\end{equation}
As a result, the kinetic energy of a beam proton follows
\begin{equation}
K=c\sqrt{p_0^2(1-z/R)^{\frac{2}{q+1}}+M^2c^2} - Mc^2 \ .
\label{eq-Kz}
\end{equation}
Figure \ref{K-vs-z} shows how this energy behaves with the choice $K_0=85\,$MeV ($p_0=408\,$MeV/c), $R=6\,$cm, and $q=1.4$ (This $q$ coming from a fit of $p=1+q/2$ performed by \citet{ulmer2010}.)

\begin{figure}[!ht]
\caption{LET: Proton energy loss per unit distance (LET)}
\center
\includegraphics[width=0.7\columnwidth]{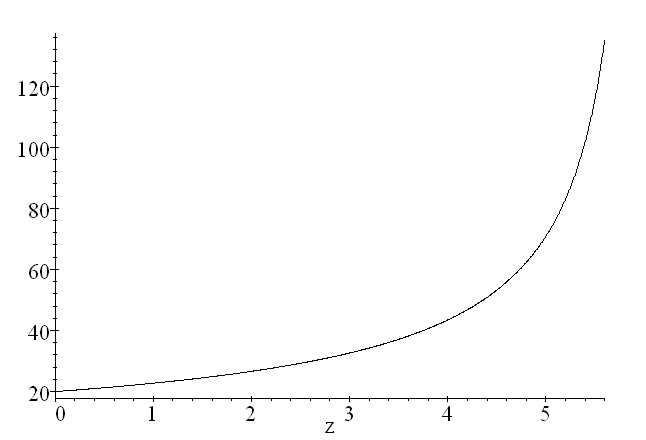}
\label{dK-vs-z}
\end{figure}

With Eq.\,(\ref{eq-Kz}), the LET of the proton as a function of penetration distance is found to be:
\begin{equation}
\frac{dK}{dz}=-\frac{c p_0^2}{(2+q)R}\frac{(1-z/R)^{-q/(q+2)}}{\sqrt{p_0^2(1-z/R)^{2/(q+2)}+\strut M^2c^2}} \ .
\end{equation}
Figure \ref{dK-vs-z} shows the corresponding slope $-dK/dz$, and exhibits the Bragg peak.

The O18 to F18 reaction occurs predominantly when $K\approx 6\,$MeV.  We can find how far ahead of the Bragg peak F81's are produced by using Eq.\,(\ref{eq-Kz}). With $K_0=85\,$MeV, the location for the O18$(p,n)$F18 nuclear reaction is greatest at about $0.6\,$mm ahead of a pristine Bragg peak set at $6\,$cm, i.e. so the lead distance is $1/100$ of $R$.

\section{Chance of beam protons hitting tissue nuclei}
\label{app-p-chance}

The interaction of a proton beam passing into a material due to scattering from material electrons and nuclei is clearly an important topic for therapeutic proton beams. Besides the energy loss
with penetration distance (as described in Sec.\ \ref{sec-K-z}), the beam is spread by multiple scattering. A successful and widely used model describing the spreading and multiple scattering of the beam began with the work of 
\citet{moliere1948}. The model gives the particle beam distribution as a function of the angle away
from the initial beam direction.
Hans Bethe \cite{bethe1953moliere} simplified the derivation. Further refinements have been
applied since 1953, including models for the relevant cross sections.  These models can be compared
to brute force Monte Carlo calculations, which can incorporate all the physics describing
a beam particle multiply scattering through even an inhomogeneous material. For a review of
such models, see Gottschalk \cite{gottschalk2018}. Since the Monte Carlo calculations take much longer run times
than current good models, proton beam planning typically uses such models.

\subsection*{Chance for a single proton to make a given number of hits to distributed nuclei}

We want to estimate the probability that a proton in passing through a
given thickness of tissue will undergo a number of nuclear-scattering events.
Let $\rho_N $ be the number density of nuclei in the tissue
and $\Delta z$ the thickness through which the protons pass. Then the
areal density of scatterers will be $\rho_N \Delta z.$ Let $A_B$ be the area
that the beam covers and $\sigma_N$ be the nuclear scattering cross-section.
From a beam-proton's perspective,
if the areas around each nucleus the size of their cross sections do not overlap, then
the probability of a proton hitting within one of the scattering-center areas is
a ratio of all the center areas to the total area covered by the proton beam, i.e. 
$P=\rho_N A_B\Delta z\,\sigma_N/A_B=\rho_N \sigma_N \Delta z\,.$
(In the frame of relativistic beam protons, the thickness $\Delta z$ measures
contracted, by a factor $\sqrt{1-(v/c)^2}=m_pc^2/E$, so the scattering-center
areas act as if they were all in a thin plane.)

With a thick target, there will be a good chance that several nuclear-cross
sections overlap when viewed from the beam-proton frame of reference.
To estimate the overlaps, let $A$ represent the average area
which covers one atom in the tissue (`atomic areas'). Divide that area $A$ into $N\sim A/\sigma $ `nuclear areas'. This number
we expect to be quite large, as nuclear cross-sections are usually less than
a barn ($(10^{-13}$cm$)^2$) while the atoms in tissue are most often separated by
$1$ to $10$ {\AA}ngstroms, making $N$ of the order $10^{10}$ or larger.
The number of scattering nuclei behind one atomic area $A$ will be $\rho_N\,A
\Delta z.$ With a fluence of $10^9$ protons/cm$^2 = 10^{-7}/${\AA}ngstrom$^2$,
the number of protons passing through an atomic area at
any given time is quite small. 

\begin{figure}[thpb]
\caption{Scattering tube for three atoms aligned with the proton beam. The atomic nuclei are projected onto the face of the
tube. A beam proton happens to pass through one of the nuclear cross sections. The configuration
of the depicted scattering centers we represented by a vector $\{1,1,1,0,\cdots,0\}$, here of dimension
$N=8^2=64$.}
\center
\includegraphics[width=0.4\columnwidth]{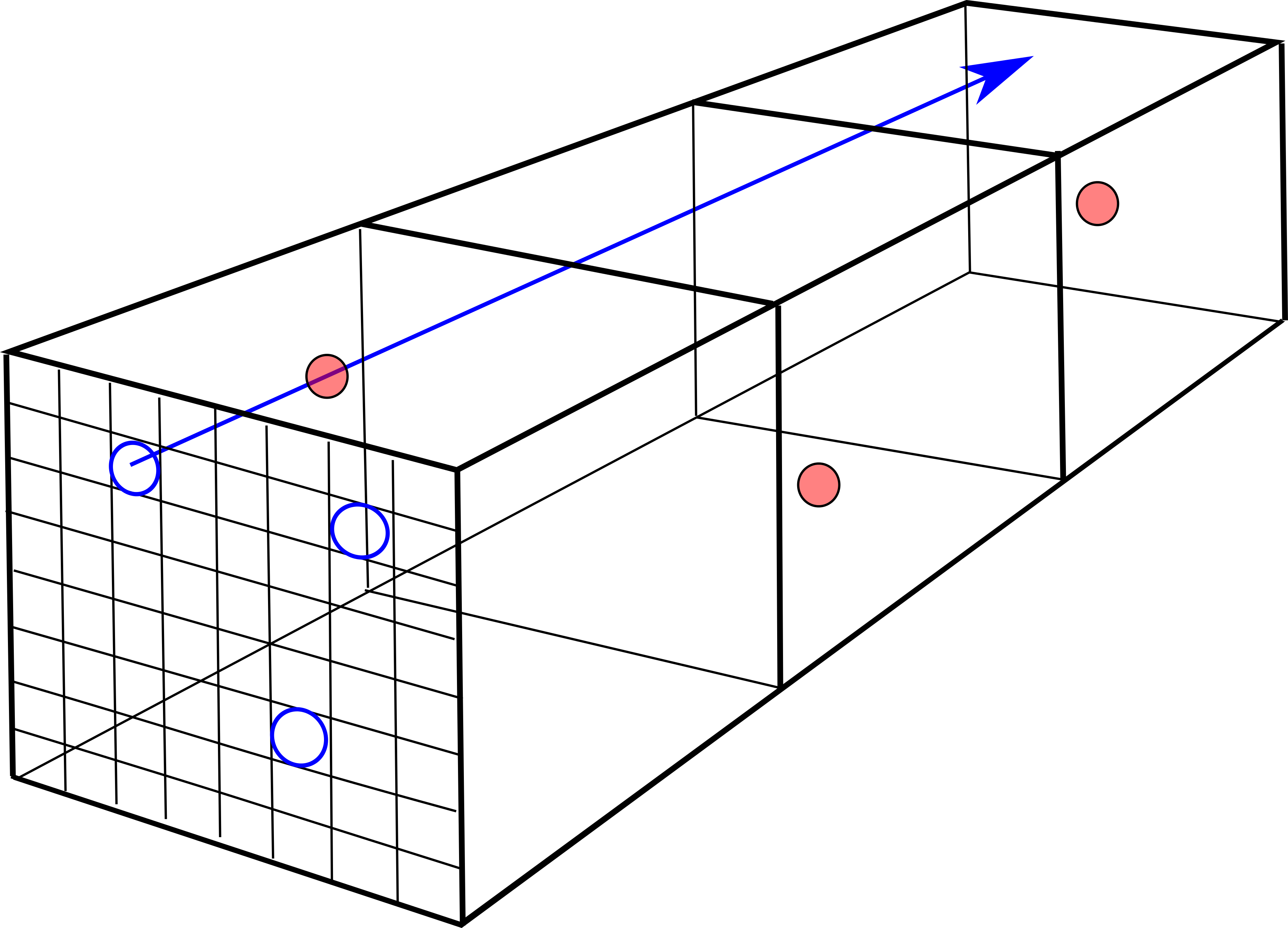}
\label{fig-p-tube}
\end{figure}

We will assume no correlation in position of the scattering centers.
A crystal will not have this property, but an amorphous mixture will, if
thick enough. As a beam proton passes through 
an atomic area $A$, while traversing a distance $\Delta z$, the atomic volumes
it passes through will make up a `scattering tube' surrounding the path of the beam proton. The number 
of atomic volumes and therefore the number of nuclei in the tube will be $n\sim \Delta z/A^{1/3}$.  
We will assume that the $n$ scattering centers in the scattering tube are
spread randomly, so that the beam proton entering any of the $N$
nuclear areas has the same probability of being scattered as
a proton entering any other such nuclear area.

Given the equal a priori probabilities on entering any one of the nuclear areas,
we can arrange all the nuclear areas covering the two dimensional atomic
area $A$ into a linear array of `cells', symbolized by an $N$ dimensional
vector.   Behind one of those
nuclear areas there may be $k>0$ nuclear scattering centers, which we
will call a cluster. The number of distinct clusters of scattering centers
distributed across cells will be called $c$.  Because
the position of the nuclear areas does not change the a priori probability
for scattering, we can order the cell `occupation' numbers $k$
so that, from left to right, $k_1\ge k_2 \ge k_3 \cdots \ge k_c$.

A single configuration for scattering centers will be called a `distribution' of fixed values for the $k_i$,
with the distribution `vector' symbolized by an $N$ dimensional vector
\begin{equation}
\vec{k}\equiv \{k_1,k_2,\cdots,k_c,0,0,\cdots\}= \{k_1,k_2,\cdots,k_N\} \ .
\end{equation}
In the second expression, the $k_i$ values beyond $i=c$ are all zero.
Note that
\begin{equation}
\sum_{i=1}^N k_i=n \ .
\end{equation}
Evidently, the number of clusters is
\begin{equation}
c\equiv N-\text{number of empty cells} \ 
\end{equation}
and $1\le c \le \min{(n,N)}$.
A distribution vector is then $\{k_1,k_2,\cdots,k_c,0,0,\cdots,0\}$. 
If a subset of non-zero $k$'s
have the same value, we will call the number of such equal $k$'s the `multiplicity'$\equiv m$ of that $k$. 

After randomly distributing the $n$ scattering centers in  $N$ cells, the number of configurations of scattering-center clusters with a given distribution
$\{k_1,k_2,k_3,\cdots,k_N\}$ will be called the weight of the distribution, $N_c(\{k\})$, since this
number is proportional to the probability of finding such a configuration after the
scattering centers are randomly distributed among the $N$ cells.

\begin{figure}[thpb]
\caption{Hand-calculation of number of configurations of  five scattering centers in three cells.  A single center (one `dot') is distributed first, and then probabilistic equivalence is used to represent
the three possibilities as three equivalent ones like the first. In the figure for the distributions with
four and five scattering centers, only the resulting configurations and their count are shown.}
\center
\includegraphics[width=0.4\columnwidth]{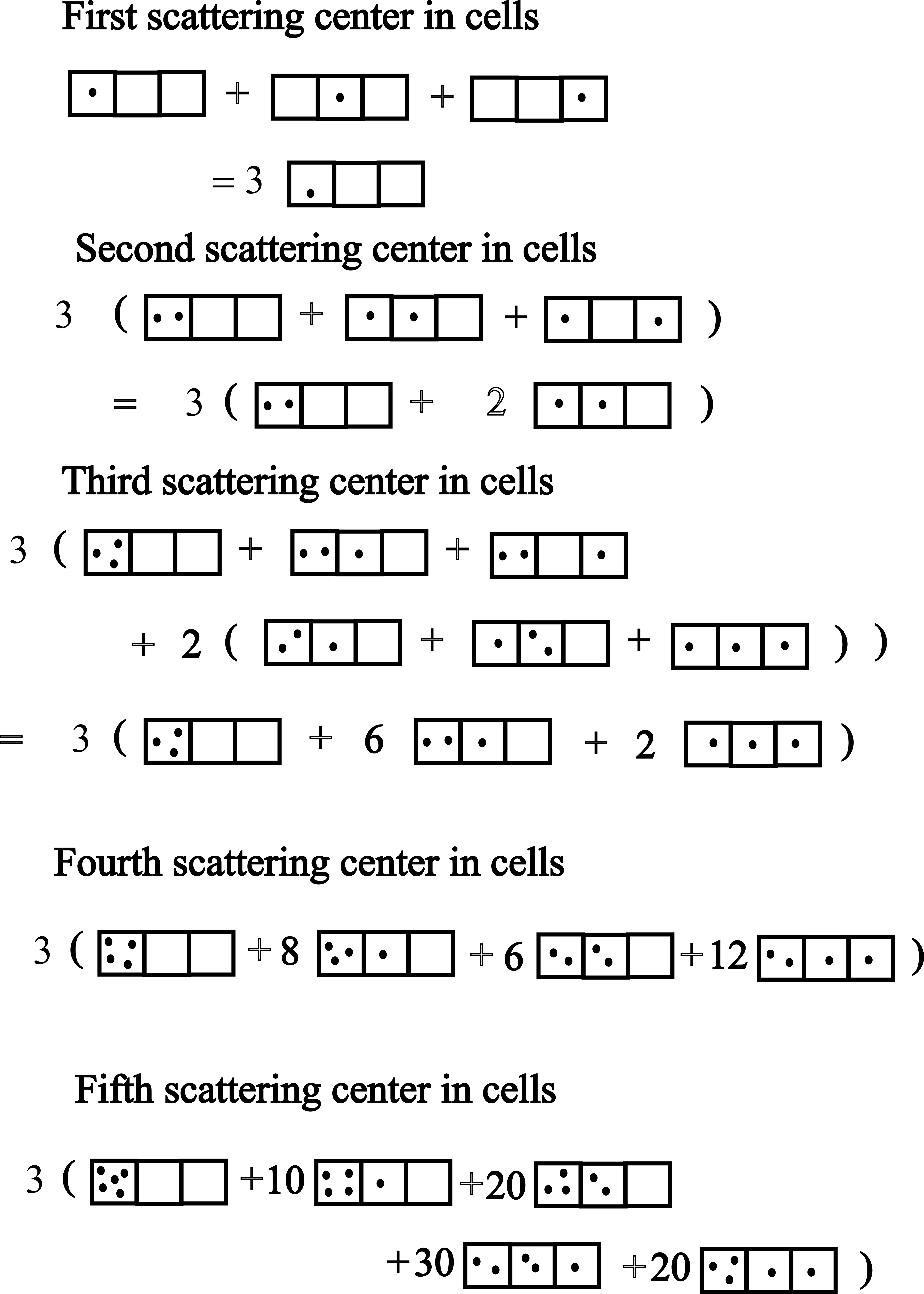}
\label{fig-cells-Nc}
\end{figure}
Figure \ref{fig-cells-Nc} shows  the case for $N=3,\ \ n=5$.  By filling three cells sequentially with five `dots' and then counting, we find the following configurations and their probabilistic weights $N_c$:
\begin{center}
$
\begin{array}{cccccccr}
$config$  && c && m_1 & m_2  && N_c\\
\{5,0,0\} && 1 && 1   &      &&  3 \\
\{4,1,0\} && 2 && 1   &  1   && 30 \\
\{3,2,0\} && 2 && 1   &  1   && 60 \\
\{2,2,1\} && 3 && 2   &  1   && 90 \\
\{3,1,1\} && 3 && 1   &  2   && 60 
\end{array} 
$
\end{center}
In this example, the total number of configurations must be $N^n=3^5=243$, a useful check that all configurations have been found.

To find a general
expression for the weights $N_c(\{k\})$, first note that if there are $c$ clusters, 
there will be $N(N-1)\cdots(N-c+1)=N!/(N-c)!$ ways to distribute 
distinct clusters.  If there are $m_l$ cells with the same number $k_l>0$
of scattering centers, then the number of configurations with 
a given set of clusters is $N!/((N-c)!\prod m_l!)$, since
a cluster with label $l$ has $m_l!$ ways of equivalent re-arrangements.
Now we must count the number of ways the scattering centers could have
been placed in the given clusters. Starting with the left-most non-empty
cluster which contains $k_1$ scattering centers, there will have been $\binom{n}{n-k_1}$
ways to have selected the centers. But now we have $n-k_1$ fewer centers 
to put in the second cluster, so there will have been $\binom{n-k_1}{n-k_1-k_2}$
ways to rearrange the centers in the second cluster, given the first. This continues
until the last non-zero cluster, which has $\binom{n-\sum_{i=1}^{c-1}k_i}{k_c}=\binom{k_c}{k_c}=1$.
As a result, the number of distributions with vector $\{k_1,k_2,\cdots k_N\}$
will be
\begin{equation}
N_c(\{k\})  = \frac{N!}{\left( N-c\right) !\prod m_{l}!}%
\binom{n}{k_{1}}\binom{n-k_{1}}{k_{2}}\binom{n-k_{1}-k_{2}}{k_{3}}\cdots 
\binom{n-\sum_{1}^{c-2}k_{i}}{k_{c-1}}\binom{n-\sum_{1}^{c-1}k_{i}}{k_{c}}  \ .
\end{equation}
i.e.
\begin{equation}
N_c(\{k\})=\frac{N!}{(N-c)!\prod m_l!}\frac{n!}{\prod k_i!} \ .
\label{eq-Nc}
\end{equation}
The counts $N_c(\{k\})$ must satisfy
\begin{equation}
\sum_{c}N_c(\{k\})=\sum_{\left\{ k_{1},\cdots k_{c}\right\} }\frac{N!}{%
\left( N-c\right) !\prod m_{l}!}\frac{n!}{\prod_{1}^{c}k_{i}!}=N^{n}
\label{eq-Nid}
\end{equation}
wherein the sum is over all $k_{i}$ that satisfy $k_{1}\geq
k_{2}\geq \cdots k_{c}$ and $\sum_{i=1}^{c}k_{i}=n.$ 

The identity Eq.\,(\ref{eq-Nid}) can be derived from the following
observation.  The multinomial expansion is
\begin{equation}
(a_1+a_2+a_3+\cdots+a_N)^n=
\sum_{\{k_i=0\}}^N \frac{n!}{k_1!k_2!\cdots k_N!}a_1^{k_1}a_2^{k_2}\cdots a_N^{k_N} \ ,
\end{equation}
where $\sum_{i=1}^Nk_i=n$ and now all the $k$'s range from $0$ to $N$.
If we put all the $a$'s to one, then we have 
\begin{equation}
N^n=\sum \frac{n!}{k_1!k_2!\cdots k_N!}\ .
\end{equation}
In this sum, group all terms that have the same set of $\{k_1,k_2,\cdots k_N\}$.  Consider the terms
with $c$ non-zero $k$'s. For the case of distinct $k$'s, there
will be $N(N-1)\cdots (N-c)$ such terms.  But some terms with a given $c$ may have a degeneracy, i.e.
terms with $m_l$ identical $k$'s. Then the number of distinct terms is $N(N-1)\cdots (N-c)/\prod m_l!$. 
The multinomial for a sum of ones has become the sum of the $N_c(\{k\})$. 

A useful alternative to the expression Eq.\,(\ref{eq-Nc}) is 
\begin{equation}
N_c=\frac{N!}{(N-c)!}\frac{n!}{\prod_{l=1}^{L} m_l!(\kappa_l!)^{m_l}} \ ,
\label{eq-Nca}
\end{equation}
where the $\kappa_l$ are all the distinct $k$'s, with $\kappa_1>\kappa_2\cdots >\kappa_L>0$ and
\begin{eqnarray}
\sum_{l=1}^L m_l & = & c \\
\sum_{l=1}^L m_l \kappa_l &=& n \ .
\label{eq-mc}
\end{eqnarray}

With this notation, a given configuration of scattering centers can be denoted $(\kappa_1^{m_1},\kappa_2^{m_2},\cdots,\kappa_L^{m_L})$, where $L\le c\le \min{(n,N)}$.

An even simpler expression results if we define $m_0\equiv N-c=N-\sum m_l$ for the multiplicity of
the empty cells.  Then
\begin{eqnarray}
N_c&=&\left( \begin{array}{c} N \\ m_0\cdots m_L\\ \end{array}\right)
\left(\begin{array}{c} n \\ k_1\cdots k_c\\ \end{array}\right) \ , \\
&=&\left( \begin{array}{c} N \\ m_0\cdots m_{N}\\ \end{array}\right)
\left(\begin{array}{c} n \\ k_1\cdots k_n\\ \end{array}\right) 
\label{eq-Nca2}
\end{eqnarray}
where the two parenthetical expressions are multinomial coefficients, and we have taken
advantage of the fact that $\sum_0^Lm_l=m_0+\sum_1^Lm_l=N-c+c=N$. In Eq.\ (\ref{eq-Nca2}), we assign zeros to the multiplicities $m_l$ for $L<l< N$, and then use
the fact that any number of $m$'s or $k$'s
can be appended to the array in each multinomial, as long as they are zero.

The probability for a given configuration of $\{k_1,k_2,\cdots k_c\}$ will be
\begin{equation}
P_c\{k\}=\frac{N_c}{N^n} \ .
\label{eq-pck}
\end{equation}

Evidently, the configurations with the scattering centers spread out (having the $k$'s close to the
same values) will have the larger probabilities, but, as seen in the example for $N=5,\ n=7$ below, a configuration
with fewer clusters can have a larger probability than one with a more evenly
spread $k$. (This is in contrast to the multinomial coefficients themselves, for which a more even
spread of the $k$'s always gives a larger coefficient.)

The weight for the probability $P_{>1}$ that two or more scattering events (`hits') occur for a given beam proton will be a sum of the weight of a given cluster set times the probability $p_{>1}$ of a hit in a cluster whose size $k_i$ is greater than one. This determination might best be seen in an example. Let $N=5$ and $n=7$. Then the possible clusters with their
weights $N_c$ are shown in Table \ref{tab-confs}.
\begin{table}[!htbp]
\caption{Partitions of 7 centers in 5 cells, with the number of each partition given by $N_c$ and the probability of a hit on a cell with more than one center given by $P_{>1}$.}
\begin{center}
$
\begin{array}{ccccrrrr}
\left\{ k_1,k_2,\cdots\right\} &  & & N_c &\times & p_{>1}   \\[2pt]
\hline
\left\{ 7,0,0,0,0\right\}  & 
&  & 5 & \times & 1/5 &=& 1\\[4pt]
\left\{ 6,1,0,0,0\right\}  &  
&  & 140 & \times & 1/5 &=& 28 \\[4pt]
\left\{ 5,2,0,0,0\right\}   & 
&  & 420 & \times & 2/5 &=&168\\[4pt]
\left\{ 5,1,1,0,0\right\}  &  
&  & 1260 & \times& 1/5 &=&252\\[4pt]
\left\{ 4,3,0,0,0\right\}  & 
& & 700 & \times &  2/5 &=& 280\\[4pt]
\left\{ 4,2,1,0,0\right\}  &  
&  & 6300 & \times & 2/5 &=& 2520\\[4pt]
\left\{ 4,1,1,1,0\right\}  &  
&  & 4200 & \times& 1/5 &=& 840\\[4pt] 
\left( 3,3,1,0,0\right)  &  
&  & 4200 & \times& 2/5 &=& 1680 \\[4pt]
\left\{ 3,2,2,0,0\right\}  &  &  & 6300 & \times&  3/5  &=& 3780 \\[4pt]
\left\{ 3,2,1,1,0\right\}  &  &  & 25200 & \times & 2/5 &=& 10080\\[4pt]
\left\{ 3,1,1,1,1\right\}  &  &  & 4200 & \times& 1/5 &=& 840\\[4pt]
\left( 2,2,2,1,0\right)  &  &  & 12600 & \times& 3/5 &=& 7560\\[4pt]
\left( 2,2,1,1,1\right)  &  &  & 12600 & \times & 2/5 &=& 5040 \\[4pt]
sum & &  &  5^7=78125 & & && 33069
\end{array}
$
\end{center}
\label{tab-confs}
\end{table}
For this example, the fraction $P_{>1}$ of beam protons that hit two or more scattering centers
is $33069/78125$, or about $42\%$.

The general expression for $P_{>1}$ is given by
\begin{equation}
P_{>1}=\sum_{clusters}\frac{1}{N^n}\frac{n_{>1}}{N}\frac{N!}{(N-c)!\prod m_l!}\frac{n!}{\prod k_i!}
\label{eq-pc1}
\end{equation}
Here, $n_{>1}\le c$ is the number of $k$'s bigger than one in the given distribution.
Note that we have not required that the proton be deflected only to small angles.  After each hit, the
distribution of scattering centers is the same as that presented to the proton in the prior hit. This is a property
following from the implicit assumption of homogeneity and isotropy of the tissue and of random distribution of
scattering centers with cross-section $\sigma$ among the possible scattering tubes.

A general expression for $P_{\geq 1}$ is given by
\begin{equation}
P_{\geq 1}=\frac{1}{N^n}\sum_{clusters}\frac{n_{\geq1}}{N}\frac{N!}{m_0!\prod_{m_l=1}^N m_l!}\frac{n!}{\prod_{k_i=1}^n k_i!} \ .
\label{eq-P1}
\end{equation}
Here, $n_{\geq 1}\le c$ is the number of $k$'s bigger than zero in the given cluster.
Note that we have not required that the proton be deflected only to small angles.  After each hit, the
distribution of scattering centers is the same as that presented to the proton in the prior hit. This is a property
following from the implicit assumption of homogeneity and isotropy of the tissue and of random distribution of
scattering centers with cross-section $\sigma$ among the possible scattering tubes.

The expression Eq.\ (\ref{eq-P1}) can be greatly simplified. First note that $n_{\geq 1}$ is
also the sum of the multiplicities $m_l$ except for $m_0$, and that 
\[
m_0+\sum_{l=1}^nm_l=N
\]
so that 
\[
n_{\ge 1}=N-m_0 \ .
\]
Substituting into Eq.\ (\ref{eq-P1}) gives
\begin{equation}
P_{\geq 1}=\frac{1}{N^{n}}\left(\sum_{clusters}\frac{N!}{m_0!\prod_{m_l=1}^N m_l!}\frac{n!}{\prod_{k_i=1}^n k_i!}-\sum_{clusters}\frac{(N-1)!}{(m_0-1)!\prod_{m_l=1}^N m_l!}\frac{n!}{\prod_{k_i=1}^n k_i!}\right) \ .
\end{equation}
(The $m_0=0$ case is allowed in the second sum because $1/((-1)!)=0.$) We recognize each cluster sum to be an expansion of a sum of ones to a power of $n$. Thus
\begin{equation}
P_{\geq 1}=\frac{1}{N^{n}}\left(N^n-(N-1)^n\right) = 1-(1-1/N)^n \approx 1-\exp{(-n/N)}\ ,\\
\label{eq-P1a}
\end{equation}
where, in the last expression, we took $N>n>>1$.  Alternatively, this exponential behavior 
can be derived (as in Beer's law for light scattering) from the assumption that the number of scattered protons is proportional
to the number arriving into a given small volume of tissue and the density of scattering centers
in that volume.

To find the probability that a beam proton independently hits at least two scattering centers, we turn to
\begin{equation}
P_{\geq 2}=
\frac{1}{N^n}\sum_{clusters}\frac{n_{\geq 2}}{N}\frac{N!}{m_0!\prod_{m_l=1}^N m_l!}\frac{n!}{\prod_{k_i=1}^n k_i!} \ .  
\label{eq-P2}
\end{equation}
Now $n_{\geq 2}=N-m_0-m_1$. The evaluation of our expression Eq.\ (\ref{eq-P2}) is
easier as a single multinomial sum.  There results

\begin{equation}
P_{\geq 2}=1-\left(1-\frac{1}{N}\right)^n-\frac{1}{N^{n+1}}nN\sum_{k_2,...,k_N}\left(\begin{array}{c} n-1 \\ k_2\cdots k_N \ , \\ \end{array}\right) \ ,
\label{eq-P2a}
\end{equation}
where the $k_1$ sum has been excluded. The factor $N$ in front of the sum comes from the
fact that there are $N$ equivalent such sums. The ranges of the remaining $k$'s go from $0$ to $n-1$, and
they are constrained by $\sum_{i=2}^N k_i=n-1$. The sum in Eq.\ (\ref{eq-P2a}) is just $(N-1)^{n-1}$, resulting in
\begin{equation}
P_{\geq 2}=1-\left(1+\frac{n-1}{N}\right)\left(1-\frac{1}{N}\right)^{n-1}
\approx 1-\left(1+\frac{n-1}{N}\right)\exp{(-(n-1)/N)}  \ .
\label{eq-P2b}
\end{equation}

Note that if $n=N$ (number of scattering centers is the same as the number of nuclear cells),
then, for large $n$, $P_{\geq 1}\rightarrow 1-1/e=0.632$ and $P_{\geq 2}=1-2/e=0.264$.  These
serve as a check of calculations.

A Mathematica program is given in Appendix \ref{app-p-prob} which finds all
the ordered distributions $\{k_1,k_2,\cdots,k_N\}$, calculates $N_c$ for
a given distribution and then
the probabilities $P_{>0},\ P_{>1},\ \cdots$ for a beam proton to be scattered once, twice, or any a number of times.

\subsection*{Chance for a proton to undergo hits to O18 in tissue}

In our applications, the number of nuclear scattering centers 
behind the area $A$ is $\rho_NA\Delta z$, while the number of cells over which the
scattering centers are distributed in $N=A/\sigma$, so $n/N=\rho_N\sigma \Delta z.$

As the proton beam slows, the events O18(p,n)F18 increase as the reaction cross section peaks at about 6 MeV (see Fig.\,\ref{fig-O18F18-cross}). For a low density of parent nuclei, the probability that a proton will hit the cross-sectional area $\sigma$ is given by $\mathcal{P}\approx(\Delta N/\Delta V)\sigma \Delta z$, where $(\Delta N/\Delta V\equiv \rho_N)$ is the number density of the parent nuclei and $\Delta z$ is the distance the proton has traveled. When the area $N\sigma$ becomes a significant fraction of the area $A_B=\Delta V/\Delta z$, multiple independent hits are likely. According to Eq.\ (\ref{eq-P1a}), the probability of at least two independent hits in the case $1<<n<<N$ has the leading term 
$P_{\geq 2}\approx \frac{1}{2}\left(\frac{n-1}{N}\right)^2 \ ,$
while $P_{\geq 1}\approx \frac{n}{N} \ .$ 
The approximate expression for $P_{\geq 1}$ would also follow from the assumption that the change in the flux of beam particles
over a short distance drops in 
proportion to the flux itself, to the scatterer's cross section and to their number density.
The approximate expression for $P_{\geq 2}$ would come about if the change of flux over a short distance dropped as the square of the distance the beam proton traveled, indicating
that double scattering is required, just as the chance that a car is involved in two
independent encounters with other cars is proportional to the density of cars squared.

Table \ref{tab-O18-data} shows the relevant data for thymine-18 in the DNA of human tissue. One can see from the long `extinction length' $\lambda=1/(\rho_N\sigma)$ for the O18(p,n)F18 reaction that inside a human, a beam proton is not likely
to have more than one encounter with O18 to make F18, assuming the O18 is uniformly
spread out. However, the fact that the DNA is compacted into a small volume inside each biologic cell
increases the chance that one encounter will be followed by a second, i.e. scattering events
may be correlated.

For scattering of beam protons off nuclei, we expect that the number of F18 isotopes that the proton beam produces in the doped tissue will be
\begin{equation}
N_{F18}=J A_B \Delta t\left(1-\exp{(-\rho_N\sigma \Delta z)}\right)\approx
J A_B \rho_N\sigma \Delta z \Delta t \ ,
\end{equation}
where $J$ is the number flux of protons in the beam, $A_B$ is the effective area of the beam, and $\Delta t$ is the exposure time. 

\section{Beam fluence  loss}

The loss of protons from a proton-therapy beam while it passes through tissue has a number of distinct causes:
\begin{itemize}
\item Protons scattering from electrons (minimal loss)
\item Protons elastic scattering from other protons  
\item Protons inelastically scattering from nuclei making excited nuclear states 
\item Protons causing nuclear reactions (transmutation and fragmentation)
\item Protons becoming thermalized
\end{itemize}
The loss is usually measured by the change in the beam fluence rate, or number flux, defined as the number of protons passing through a unit area perpendicular to the beam per unit time. In relativity, this is the spatial part of the proton 4-current divided by the magnitude of the electron charge, i.e. the proton number density measured moving with the beam, times the relativistic 4-velocity ($J^{\mu}=\rho_o dx^{\mu}d\tau)$. 

Given the myriad of possible causes for diminishing proton flux as the beam
passed through tissue, analytic expressions for this loss as a function of
penetration distance are hard to find.  Rather, Monte Carlo techniques (e.g. GEANT4 code)
are often employed.  Even so, modeling and experiment show that the drop in proton flux with distance is almost linear.
The drop in proton number flux in water is $15\,$\% before the end of the proton beam's average range,
where the flux tails off to zero within about $3\,$\% of the beam's range,
(\cite{sandison2000}, \cite{newhauser2015}).

\section{Nuclear positron emitters produced by a proton beam}
\label{app-exam}

Reactions that produce positron-emitting radionuclides potentially could interfere with post-PT PET scans to determine where the beam delivered its dose. It behooves us to look for the reactions that
might produce significant amounts of positron nuclear decays in the time-frame of the measurable F18 decays.

We will use $(A,Z)$, to represent a nuclei.  A proton is then
represented by  $(A,Z)=(1,1)\,$, while a neutron is $(1,0)$.
After a nuclear reaction of a beam proton with a tissue nucleus, the dominant fragments will be $p,n,(pp),(pn),(nn),\alpha$.  These will be denoted
$(1,1),(1,0),(2,2),(2,1),(2,0),(4,2)$. (Deuterons, tritium, and helium-3 will have a much lower probability, as these are clustered in heavier nuclei far less often than helium-4.)
Thus, the fragments will have $\left(b,q\right)$ where $b=1,2$; $q=0,...,b$, $b=4,q=2$.

Reactions are represented succinctly by
\[(1,z)+(A,Z)=(A^{\ast },Z^{\ast })+(b,q)\] with $z=1,0$ for reactant protons or secondary neutrons.
To find the reactant nuclei for a given radionuclide, we use
\[\left( A,Z\right) =\left( A^{\ast }+b-1,Z^{\ast }+q-z\right)\]

The possible positron-emitting radionuclides that could interfere, and their half-lives, are
\begin{center}
$
\begin{array}{cc}
\hline\hline
\beta^+\text{emitter} & \text{half-life}\\
\hline
$C\,$11=(11,6) & 20.334\,$min$  \\ 
$N\,$13=(13,7) & 9.965\,$min$  \\ 
$O\,$15=(15,8) & 2.0373\,$min$  \\ 
$F\,$18=(18,9) & 1.8295\,$hrs$ \\ 
$Na\,$22=(22,11) & 2.602\,$yrs$ \\ 
$Cu\,$64=(64,29) & 12.7\,$min$  \\ 
$Ga\,$68=(68,31) & 1.1285\,$hrs$ \\ 
$Br\,$78=(78,35) & 6.46\,$min$  \\ 
$Rb\,$82=(82,37) & 1.273\,$min$  \\ 
$Sr\,$83=(83,38) & 32.41\,$hrs$ \\ 
$Y\,$86=(86,39) & 14.74\,$hrs$ \\ 
$Zr\,$89=(89,40) & 78.41\,$hrs$ \\ 
$I\,$124=(124,53) & 4.176\,$days$\\
\hline
\end{array}
$
\end{center}

\bigskip

Examinining each one up to copper, we have

\bigskip 

$C11=(11,6)\,A^{\ast }=11,Z^{\ast }=6$ \ half-life $20.334\,\min $

\bigskip 

beam proton $(z=1)$

$
\begin{array}{ccccccccc}
z & b & q &  & 
\begin{array}{c}
A= \\ 
11+b-1
\end{array}
& 
\begin{array}{c}
Z= \\ 
6+q-z
\end{array}
& name & 
\begin{array}{c}
abundance \\ 
((percent))
\end{array}
& reaction \\ 
1 & 1 & 0 & n & 11 & 5 & B & 0.0 &  \\ 
1 & 1 & 1 & p & 11 & 6 & C & 0.0 &  \\ 
1 & 2 & 0 & nn & 12 & 5 & B & 0.0 &  \\ 
1 & 2 & 1 & pn & 12 & 6 & C & 98.93 & _{6}^{12}C+p\rightarrow
\,_{6}^{11}C+n+p \\ 
1 & 2 & 2 & pp & 12 & 7 & N & 0.0 &  \\ 
1 & 4 & 2 & \alpha  & 14 & 7 & N & 99.632 & _{7}^{14}N+p\rightarrow
\,_{6}^{11}C+\alpha 
\end{array}
$

\bigskip

`beam' neutron $(z=0)$

$
\begin{array}{ccccccccc}
z & b & q &  & 
\begin{array}{c}
A= \\ 
11+b-1
\end{array}
& 
\begin{array}{c}
Z= \\ 
6+q-z
\end{array}
& name & 
\begin{array}{c}
abundance \\ 
(percent)
\end{array}
& reaction \\ 
0 & 1 & 0 & n & 11 & 6 & C & 0.0 &  \\ 
0 & 1 & 1 & p & 11 & 7 & N & 0.0 &  \\ 
0 & 2 & 0 & nn & 12 & 6 & C & 98.93 & _{6}^{12}C+n\rightarrow
\,_{6}^{11}C+n+n \\ 
0 & 2 & 1 & pn & 12 & 7 & N & 0.0 &  \\ 
0 & 2 & 2 & pp & 12 & 8 & O & 0.0 &  \\ 
0 & 4 & 2 & \alpha  & 14 & 8 & O & 0.0 & 
\end{array}
$

\bigskip 

\bigskip

$\bigskip N13=(13,7)\,A^{\ast }=13,Z^{\ast }=7$ \ half-life $9.965\,\min $

beam proton $(z=1)$

$
\begin{array}{ccccccccc}
z & b & q &  & 
\begin{array}{c}
A= \\ 
13+b-1
\end{array}
& 
\begin{array}{c}
Z= \\ 
7+q-z
\end{array}
& name & 
\begin{array}{c}
abundance \\ 
(percent)
\end{array}
& reaction \\ 
1 & 1 & 0 & n & 13 & 6 & C & 0.0 &  \\ 
1 & 1 & 1 & p & 13 & 7 & N & 0.0 &  \\ 
1 & 2 & 0 & nn & 14 & 6 & C & 0.0 &  \\ 
1 & 2 & 1 & pn & 14 & 7 & N & 99.632 & _{7}^{14}N+p\rightarrow
\,_{7}^{13}N+n+p \\ 
1 & 2 & 2 & pp & 14 & 8 & O & 0.0 &  \\ 
1 & 4 & 2 & \alpha  & 16 & 8 & O & 99.757 & _{8}^{16}O+p\rightarrow
\,_{7}^{13}N+\alpha 
\end{array}
$

\bigskip 

`beam' neutron $(z=0)$

$
\begin{array}{ccccccccc}
z & b & q &  & 
\begin{array}{c}
A= \\ 
13+b-1
\end{array}
& 
\begin{array}{c}
Z= \\ 
7+q-z
\end{array}
& name & 
\begin{array}{c}
abundance \\ 
(percent)
\end{array}
& reaction \\ 
0 & 1 & 0 & n & 13 & 5 & B & 0.0 &  \\ 
0 & 1 & 1 & p & 13 & 6 & C & 1.07 &  \\ 
0 & 2 & 0 & nn & 14 & 7 & N & 99.632 & _{7}^{14}N+n\rightarrow
\,_{7}^{13}N+n+n \\ 
0 & 2 & 1 & pn & 14 & 8 & O & 0.0 &  \\ 
0 & 2 & 2 & pp & 14 & 9 & F & 0.0 &  \\ 
0 & 4 & 2 & \alpha  & 16 & 9 & F & 0.0 & 
\end{array}
$

\bigskip \bigskip 

$O15=(15,8)\,A^{\ast }=15,Z^{\ast }=8$ half-life $2.0373\,\min $

\bigskip 

beam proton $(z=1)$

$
\begin{array}{ccccccccc}
z & b & q &  & 
\begin{array}{c}
A= \\ 
15+b-1
\end{array}
& 
\begin{array}{c}
Z= \\ 
8+q-z
\end{array}
& name & 
\begin{array}{c}
abundance \\ 
(percent)
\end{array}
& reaction \\ 
1 & 1 & 0 & n & 15 & 7 & N & 0.368 &  \\ 
1 & 1 & 1 & p & 15 & 8 & O & 0.0 &  \\ 
1 & 2 & 0 & nn & 16 & 7 & N & 0.0 &  \\ 
1 & 2 & 1 & pn & 16 & 8 & O & 99.757 & _{7}^{16}O+p\rightarrow
\,_{7}^{15}O+n+p \\ 
1 & 2 & 2 & pp & 16 & 9 & F & 0.0 &  \\ 
1 & 4 & 2 & \alpha  & 18 & 9 & F & 0.0 & 
\end{array}
$

\bigskip 

`beam' neutron $(z=0)$

$
\begin{array}{ccccccccc}
z & b & q &  & 
\begin{array}{c}
A= \\ 
15+b-1
\end{array}
& 
\begin{array}{c}
Z= \\ 
8+q-z
\end{array}
& name & 
\begin{array}{c}
abundance \\ 
(percent)
\end{array}
& reaction \\ 
0 & 1 & 0 & n & 15 & 8 & O & 0.0 &  \\ 
0 & 1 & 1 & p & 15 & 9 & F & 0.0 &  \\ 
0 & 2 & 0 & nn & 16 & 8 & O & 99.757 & _{8}^{16}O+n\rightarrow
\,_{8}^{15}O+n+n \\ 
0 & 2 & 1 & pn & 16 & 9 & F & 0.0 &  \\ 
0 & 2 & 2 & pp & 16 & 10 & Ne & 0.0 &  \\ 
0 & 4 & 2 & \alpha  & 18 & 10 & Ne & 0.0 & 
\end{array}
$

\bigskip 

\bigskip

$F18=(18,9)\,A^{\ast }=18,Z^{\ast }=9$ \ half-life $1.8295\,hrs$

\bigskip 

beam proton $(z=1)$

$
\begin{array}{ccccccccc}
z & b & q &  & 
\begin{array}{c}
A= \\ 
18+b-1
\end{array}
& 
\begin{array}{c}
Z= \\ 
9+q-z
\end{array}
& name & 
\begin{array}{c}
abundance \\ 
(percent)
\end{array}
& reaction \\ 
1 & 1 & 0 & n & 18 & 8 & O & 0.0 &  \\ 
1 & 1 & 1 & p & 18 & 9 & F & 0.0 &  \\ 
1 & 2 & 0 & nn & 19 & 8 & O & 0.0 &  \\ 
1 & 2 & 1 & pn & 19 & 9 & F & 100. & _{9}^{19}F+p\rightarrow \,_{9}^{18}F+n+p
\\ 
1 & 2 & 2 & pp & 19 & 10 & Ne & 0.0 &  \\ 
1 & 4 & 2 & \alpha  & 21 & 10 & Ne & 0.27 & 
\end{array}
$

\bigskip 

`beam' neutron $(z=0)$

$
\begin{array}{ccccccccc}
z & b & q &  & 
\begin{array}{c}
A= \\ 
18+b-1
\end{array}
& 
\begin{array}{c}
Z= \\ 
9+q-z
\end{array}
& name & 
\begin{array}{c}
abundance \\ 
(percent)
\end{array}
& reaction \\ 
0 & 1 & 0 & n & 18 & 9 & F & 0.0 &  \\ 
0 & 1 & 1 & p & 18 & 10 & Ne & 0.0 &  \\ 
0 & 2 & 0 & nn & 19 & 9 & F & 100. & _{9}^{19}F+n\rightarrow \,_{9}^{18}F+n+n
\\ 
0 & 2 & 1 & pn & 19 & 10 & Ne & 0.0 &  \\ 
0 & 2 & 2 & pp & 19 & 11 & Na & 0.0 &  \\ 
0 & 4 & 2 & \alpha  & 21 & 11 & Na & 0.0 & 
\end{array}
$

\bigskip 

\bigskip

$Na22=(22,11)\,A^{\ast }=22,Z^{\ast }=11$ \ half-life $2.602\,yrs$

\bigskip 

beam proton $(z=1)$

$
\begin{array}{ccccccccc}
z & b & q &  & 
\begin{array}{c}
A= \\ 
22+b-1
\end{array}
& 
\begin{array}{c}
Z= \\ 
11+q-z
\end{array}
& name & 
\begin{array}{c}
abundance \\ 
(percent)
\end{array}
& reaction \\ 
1 & 1 & 0 & n & 22 & 12 & Mg & 0.0 &  \\ 
1 & 1 & 1 & p & 22 & 13 & Al & 0.0 &  \\ 
1 & 2 & 0 & nn & 23 & 12 & Mg & 0.0 &  \\ 
1 & 2 & 1 & pn & 23 & 13 & Al & 0.0 &  \\ 
1 & 2 & 2 & pp & 23 & 14 & S\,i & 0.0 &  \\ 
1 & 4 & 2 & \alpha  & 25 & 14 & S\,i & 0.0 & 
\end{array}
$

\bigskip 

`beam' neutron $(z=0)$

$
\begin{array}{ccccccccc}
z & b & q &  & 
\begin{array}{c}
A= \\ 
22+b-1
\end{array}
& 
\begin{array}{c}
Z= \\ 
11+q-z
\end{array}
& name & 
\begin{array}{c}
abundance \\ 
(percent)
\end{array}
& reaction \\ 
0 & 1 & 0 & n & 22 & 11 & Na & 0.0 &  \\ 
0 & 1 & 1 & p & 22 & 12 & Mg & 0.0 &  \\ 
0 & 2 & 0 & nn & 23 & 11 & Na & 100. & _{11}^{23}Na+n\rightarrow
\,_{11}^{22}Na+n+n \\ 
0 & 2 & 1 & pn & 23 & 12 & Mg & 0.0 &  \\ 
0 & 2 & 2 & pp & 23 & 13 & Al & 0.0 &  \\ 
0 & 4 & 2 & \alpha  & 25 & 13 & Al & 0.0 & 
\end{array}
$
\vspace{0.1in}

Details about these reactions are given in Tables \ref{tab-reactions} and \ref{tab-react2}.

\section{Mathematica expressions for proton scattering probabilities}
\label{app-p-prob}

In Appendix \ref{app-p-chance}, we derived a way of calculating the probability that a beam proton
will scatter from a scattering center in tissue at least a certain number of times.  Here,
we will give Mathematica expressions to perform this kind of calculation when the number
of cells, $N$, and the number of scattering centers spread among those cells, are 
tractable by Mathematica.  This limits $N$ to be below about 100, and $n$ to be
below about $60$, so the
expressions given make a useful test-bed for examining the model, but not for cases
in which $N$ reaches billions or more.

We distribute $n$ scattering centers randomly over $N$ `cells'.  A given distribution
of scattering centers in cells is displayed with the notation $(k_1,k_2,\cdot, k_N)$,
where the $k$'s are the number of scattering centers which have ended in a given cell. Since all cells have equivalent in a priori probability, the cell distributions can be ordered according to
the size of the $k$'s as $k_1\geq k_2\cdots \geq k_N$. The $k$'s are constrained by $\sum k_i = n$.
The total
number of distinct distributions is some number $c$. For a given distribution,
some of the $k$ values may be the same. We say the $k$ scattering centers
form a `cluster'.   If so, put these clusters in groups.  Each
such group will be a collection of a certain number of cells. The number of 
such clusters we will call $m_l$,
the `multiplicity' of the $k$ value. The $m_l$ values are constrained by $\sum m_l=c$ and by
$\sum m_l k_l=n$. It is useful to extend the set of $m_l$ values with zeros, with
the multiplicity of zeros being
$m_0=N-c$. The number of 
equivalent configurations with the same $(k_1,k_2,\cdot, k_N)$ created after a random distribution of the $n$ scattering centers among the $N$ cells is called $N_c$, and was shown in Appendix \ref{app-p-chance} to be
\begin{equation}
N_c=\left( \begin{array}{c} N \\ m_0\cdots m_{N-1}\\ \end{array}\right)
\left(\begin{array}{c} n \\ k_1\cdots k_n\\ \end{array}\right) \ ,
\end{equation}
where $m_i$ is  the `multiplicity' of the cells with the same number $k$ of scattering
centers. The $m_l$ are taken once for the whole set of such cells.
In Mathematica, after setting values for variables $NN$ (for the value of $N$) and $n$, the distinct configurations can be listed using
\begin{lstlisting}
        cf = IntegerPartitions[n, {NN}, Range[0, n]] ;
\end{lstlisting}
We have put $N\rightarrow NN$ to avoid the Mathematica function $N$. A list of the values
of $N_c$ for each distinct configuration will result from
\begin{lstlisting}
        Multinomial @@@ (Tally/@ cf)[[All,All,2]]*Multinomial @@@ cf 
\end{lstlisting}
Now, in order to get the probability of at least $h$ hits by a beam proton,
we define 
\begin{lstlisting}
        pos[x_] := If[x >= h,1,0]
\end{lstlisting}
and use, sequentially, $h=0,1,2,\cdots$.
It follows that 
\begin{lstlisting}
        Tr[Flatten[Multinomial @@@ (Tally/@ cf)[[All,All,2]]*Multinomial @@@ cf*
           Map[pos,cf,{2}] ] ] / NN^(n + 1)
\end{lstlisting}
will give the probability that a beam proton will hit at least $h$ scattering centers. The implied
factor $N_c/N^n$ is the probability that a given configuration appears in the random
distribution of scattering centers, and the implied factor $n_h/N$ gives
the probability that a beam proton will hit a cell containing at least $n_h$ scattering centers.
For $n\gtrapprox 70$, the above Mathematica formula has computation times in minutes or hours
on a PC.

The number of distinct partitions of an integer $n$ taken $NN\equiv k$ at a  time is often called $p(n,k)$,
with $p(n,n)\equiv p(n)$
The number $p(n,k)$ can be calculated in Mathematica for
a given set of distinct distributions from the `length' of $cf$, i.e.
\begin{lstlisting}
        p[n_,k_):=Length[IntegerPartitions[n,k]]
\end{lstlisting}
For the example, from Table \ref{tab-confs}, with $n=7$ and $k=5$, $p(n,k)=13$.

In 1917, Hardy and Ramanujan \cite{hardy1917asymptotic,hardy1918asymptotic}
gave an asymptotic expression for $p(n)$:
\begin{equation}
p(n)\approx \frac{\pi ^2 e^{\nu} \left(\nu-1\right)}{6 \sqrt{3} \nu^3}+O\left(\frac{e^{{\nu}/{2}}}{n}\right)
\end{equation}
where $\nu=\sqrt{(2/3)\pi^2(n-1/24)}$.

Many researchers since 1917 have developed improvements to the Hardy and Ramajujan result, including convergent series for large $n$.  Here, we give that of Brassesco and Meyroneic\cite{brassesco2020}:
\begin{equation}
p(n)\approx 
\left(\frac{2 \pi ^2}{3 \sqrt{3}}\right)\frac{\exp{(r(n))}}{(2 r(n)+1)^2}\left(1-\sum_{l=1}^K\frac{D(l)}{(2 r(n)+1)^l}+O(1/n^{K/2})\right)
\end{equation}
where
\begin{eqnarray}
& D(l) \equiv & (-1)^{l+1}\frac{(l+1)}{4^l}\sum_{k=0}^{l+1}\binom{2l}{k}\frac{(-2)^k}{(l+1-k)!} \ ,\\
& r(n) \equiv & \sqrt{\frac{2}{3} \pi ^2 \left(n-\frac{1}{24}\right)+\frac{1}{4}} \ .
\end{eqnarray}
The value of the integer $K$ determines the error of the approximation.  For example, values as little $K=3$ give 
an $p(71)$ within $0.002\%$ of the exact answer, $4,697,205$.

\bigskip 

\cleardoublepage
\bibliographystyle{apalike}

\bibliography{parke-bib}

\end{document}